\documentclass[aps,pra,reprint,groupedaddress]{revtex4-2}

\usepackage[utf8]{inputenc}
\usepackage[english]{babel}
\usepackage[T1]{fontenc}
\usepackage{amsmath,amssymb,amsthm,thmtools}
\usepackage{mathtools,mathrsfs}
\usepackage{amsbsy}
\usepackage{hyperref}
\usepackage{tikz}
\usepackage{braket}
\usepackage{tikz-cd}
\usepackage{qcircuit}
\usepackage{float}
\usepackage[caption = false]{subfig}
\usepackage{color}
\usepackage{tabularx}
\usepackage{orcidlink}
\usepackage{multirow}

\newtheorem{theorem}{Theorem}

\begin{document}

\title{Efficient particle-conserving brick-wall quantum circuits}

\author{Babatunde M. Ayeni \orcidlink{0000-0003-4035-149X} }
\email{babatunde.ayeni@mu.ie}
\affiliation{Department of Theoretical Physics, National University of Ireland, Maynooth, Ireland}

\date{\today}

\begin{abstract}
In variational quantum optimization with particle-conserving quantum circuits, it is often difficult to decide \emph{a priori} which particle-conserving gates and circuit ansatzes would be most efficient for a given problem. This is important especially for noisy intermediate-scale quantum (NISQ) processors with limited resources. While this may be challenging to answer in general, deciding which particle-conserving gate would be most efficient is easier within a specified circuit ansatz. In this paper, we show how to construct efficient particle-conserving gates using some practical ideas from symmetric tensor networks. We derive different types of particle-conserving gates, including the generalized one. We numerically test the gates under the framework of brick-wall circuits. We show that the general particle-conserving gate with only four real parameters is generally best. In addition, we present an algorithm to extend brick-wall circuit with two-qubit nearest-neighbouring gates to non-nearest-neighbouring gates. We test and compare the efficiency of the circuits with Heisenberg spin chain with and without next-nearest-neighbouring interactions. 
\end{abstract}


\maketitle

\section{Introduction}
Despite the impressive advancements seen in quantum computing technology in just a few decades, current and near-term quantum processors are limited in scale and noisy, hence limiting their applications. This motivated their descriptive name of  ``noisy intermediate-scale quantum'' (NISQ) computers.\cite{preskill2018quantum} 

One important problem that is suitable for NISQ processors is calculating ground state of many-particle systems in quantum chemistry, material science, or quantum many-body physics. This generally involves building a quantum circuit to prepare a trial state, which is then optimized to the ground state of the problem with the help of some quantum algorithm. An approach that is currently feasible for NISQ devices is the variational quantum eigensolver (VQE) algorithm.\cite{peruzzo2014variational} This is a hybrid quantum-classical algorithm, where the trial state is  prepared as a parameterized quantum circuit on a quantum processor, and the optimization is performed on a classical computer.\cite{cerezo2021variational,mcclean2016theory,bharti2022noisy} 

It has been discovered that in the limit of a high number of qubits and hence exponentially large Hilbert space, VQE can suffer from ``barren plateaus.''\cite{mcclean2018barren,wang2021noise, cerezo2021cost,ragone2023unified} {These are regions of high-dimensional parameter space where the gradient becomes close to zero during a training process. This leads to slow or stalled learning, and thus rendering variational approaches ineffective. Several methods have been developed to mitigate this problem, includcing using a local rather than global cost function,\cite{cerezo2021cost} compiling deep quantum circuits into shallow ones,\cite{robertson2022escaping,robertson2023approximate} parameter initialization strategies,\cite{grant2019initialization,zhang2022escaping,wang2023trainability} and many others.\cite{grant2019initialization,sack2022avoiding,patti2021entanglement}}

It is well known that physical symmetries can be used to reduce the number of parameters needed to describe a system. This can be utilized to construct parameterized quantum circuits (PQCs) with reduced number of parameters, and hence lower computational cost. This is especially important in NISQ computing where quantum resources are limited. The idea of exploiting symmetries in quantum circuits is reviewed in Ref.~\cite{lacroix2023symmetry}. For the specific case of conserving global particle number in quantum circuits, which is also the focus of our paper, several works have been done.\cite{barkoutsos2018quantum,arrazola2022universal,gard2020efficient,yordanov2020efficient,anselmetti2021local,lee2018generalized,xie2022qubit,xia2020qubit} The circuits proposals in the papers fall into different categories of particle-conserving circuit ansatzes, depending on their areas of applications in either quantum chemistry or condensed-matter. But without an exhaustive comparison of the circuits for various applications, it is hard to justify \emph{a priori} which would be most efficient for a particular application and quantum device architecture. Moreover, the parameterization used in the circuits were chosen arbitrarily without justification. While it may be difficult to determine which circuit is generically best for all applications and still efficient for NISQ, the question of which particle-conserving gate is best is relatively cheaper to answer. In this paper, we investigate the latter question under the framework of hardware-efficient brick-wall circuits. Furthermore, many circuit ansatzes geared toward quantum chemistry applications usually include long-range gates, which may for instance be used to create single or double excitations across qubits separated by intermediate qubits. We present an algorithm to construct brick-wall circuits with long-range excitation gates. Through numerical simulations we compare the efficiency of such circuits with their counterparts consisting of only nearest-neighbouring two-qubit gates. 

One field that is very close to quantum circuit theory where the exploitation of symmetries have been well developed is tensor networks. This developments resulted in a plethora of works by many authors in around the past two decades.\cite{singh2010tensor,singh2011tensor,singh2012tensor,bauer2011implementing,mcculloch2002non,weichselbaum2012non,pfeifer2010simulation,konig2010anyonic,pfeifer2015finite,singh2014matrix,ayeni2016simulation,schmoll2020programming} Mathematically, quantum circuits can be described as tensor networks, and therefore some of the ideas used in the construction of symmetric tensors and tensor networks could find applications in symmetric quantum circuit constructions. Symmetries are exploited in simulations with quantum circuits for the same purposes for which they are exploited in tensor networks, namely, $(1)$ To reduce parameter space, thereby lowering computational cost.  $(2)$ Can be used to enforce the preservation of symmetries of the system, for example, symmetry-protected phases. 

We import some of the practical methods used in tensor networks to construct efficient quantum gates that conserve global particle number. Our construction of particle-conserving gates is based on the representation of $\mathbb{Z}_2$. We first consider two qubits that are nearest neighbouring, and later extend to any two qubits that are non-neighbouring. We realize efficient circuits for particle-conserving gates for many different types of parameterizations. The generalized particle-conserving gate is also presented. We then use the gates as ``modules''---building blocks---to construct hardware-efficient brick-wall circuits. We present algorithms for the construction of hardware-efficient brick-wall parameterized quantum circuits with nearest-neighbouring and non-nearest-neighbouring particle-conserving gates. We test the circuits on some learning problems, including Heisenberg spin chain with nearest-neighbouring (NN) and/or next-nearest-neighbouring (NNN) interactions, and learning of random quantum states. We discovered that among the particle-conserving gates, the generalized one performed best. For the circuits, we discovered that the ones with long-range particle-conserving gates with open boundary condition do not necessarily perform better than circuits consisting of only NN gates.

The rest of the paper is structured as follows: In Sec.~\ref{Sec:Symmetric circuit elements}, we present our methodology for systematically building symmetric parameterized quantum gates. We restrict our consideration to the case of $\mathbb{Z}_2$ symmetry and conserved particle number. In Sec.~\ref{Sec:PC gates}, we provide a list of inequivalent types of parameterizations  for particle-conserving two-qubit gates. In Sec.~\ref{Sec:PQCs}, the gates are used to construct particle-conserving brick-wall circuits. In Sec.~\ref{Sec:Test Model}, we performed numerical experiments with the particle-conserving brick-wall circuits, present results and discussions. In Sec.~\ref{Sec:Conclusion}, we conclude the paper.

\section{Generic $\mathbb{Z}_2$-symmetric and particle-conserving gates}
\label{Sec:Symmetric circuit elements}
By importing ideas from symmetric tensor networks, we lay out a general framework that can be used to construct efficient symmetric $\mathbb{Z}_2$ and particle-conserving quantum gates on two qubits, which may not necessarily be nearest-neighbouring on a quantum processor. In this section we first show the general idea, and later present generic circuits for symmetric gates for two or more qubits. Our approach is constructive, so not only can it be used to realize efficient particle-conserving gates on two qubits, but it can also be applied to multiple qubits. We will use (the optimal forms of) these circuits as ``modules'' to construct hardware-efficient ansatzes for variational quantum optimization.

\subsection{The general idea}
\label{Sec:General Idea}
We present the general idea for a bipartite system, but it can be extended easily to an $n$-qubit quantum system. Let $\mathcal{M}: \mathbb{V^{(A)}} \otimes \mathbb{V^{(B)}} \rightarrow \mathbb{V^{(A)}} \otimes \mathbb{V^{(B)}}$ be a unitary operation on the tensor product space $\mathbb{V^{(A)}} \otimes \mathbb{V^{(B)}}$. If $\mathcal{M}$ enjoys a  symmetry, then there exists a symmetry basis where $\mathcal{M}$ has a ``nice'' form, for instance it may be written as a block-diagonal matrix, say $Q$. There would also be basis transformations that map between the product basis and symmetry basis. Therefore, the unitary operator $\mathcal{M}$ can be obtained by composing all the parts together. The steps are outlined as 
\begin{enumerate}
    \item First, apply a map $\mathcal{F}$ to transform from the product space to a symmetry basis on a composite space, $\mathcal{F} : \mathbb{V^{(A)}} \otimes \mathbb{V^{(B)}} \rightarrow \mathbb{V^{(AB)}}$. In tensor networks $\mathcal{F}$ is also called a ``fusion''  map.

    \item Then construct a charge-conserving operator $\mathcal{Q}$ on the composite space $\mathbb{V^{(AB)}}$ in the symmetry basis.

    \item Finally, apply an inverse map $\mathcal{S}$ to transform back to the product space from the composite space, i.e. $\mathcal{S} : \mathbb{V^{(AB)}} \rightarrow \mathbb{V^{(A)}} \otimes \mathbb{V^{(B)}}$. In tensor networks $\mathcal{S}$ is also called a ``splitting''  map. 
\end{enumerate}
All the operations are then composed together to realize
\begin{equation}
\label{Eq:Symmetric M composition}
    \mathcal{M} = \mathcal{S} \circ \mathcal{Q} \circ \mathcal{F},
\end{equation}
which can be summarized with the commutative diagram 
\begin{equation}
\begin{matrix}
\begin{tikzcd}[sep=huge]
\mathbb{V}^{(A)}  \otimes \mathbb{V}^{(B)}  \arrow[r,shift left,"\mathcal{M}"]  \arrow[d,shift left, "\mathcal{F}"]  & \mathbb{V}^{(A)}  \otimes \mathbb{V}^{(B)} \\
\mathbb{V}^{(AB)} \arrow[r,shift left,"\mathcal{Q}"] & \mathbb{V}^{(AB)} \arrow[u,shift left,"\mathcal{S}"]
\end{tikzcd}
\end{matrix}.
\end{equation}
The remaining job is to find the circuits for each part of the product, and compose them together. Equivalently also, the matrix representation $M$ can be obtained easily if the matrix representation $F$, $Q$, and $S$ of operators $\mathcal{F}$, $\mathcal{Q}$, and $\mathcal{S}$ are known.

In the next sections, we show that while charge-conserving operators in a symmetry basis can be realized as controlled operations in general,  the circuits for the ``fusion'' and ``splitting'' basis transformations will depend on the symmetry group in question. Our focus will be only on operators respecting $\mathbb{Z}_2$ and particle number symmetries, and the basis transformations for those cases are easily obtainable. 

We shall examine each part of the above commutative diagram individually, starting with the charge-conserving operator $\mathcal{Q}$.

\subsection{Charge-conserving operator in a symmetry basis}
\label{Sec:Symmetric Op quantum circuit}
An operator that respects the symmetry of a quantum system can be written in the symmetry's basis. In matrix representation, the operator may be written as a block-diagonal matrix, where each block is associated with a conserved charge. Here, we present a general way to construct quantum circuits for charge-conserving operators in a symmetry basis. To achieve this, we shall make use of the following proposition.

\begin{theorem}
Charge-conserving operators in a  symmetry basis can be realized as controlled operations, and vice versa.
\label{Thm: Charge-conserving as controlled operators}
\end{theorem}

We ``prove'' this statement in two directions. In the first, we deal with the first part of the statement: ``\emph{charge-conserving operators can be implemented as controlled operations}.'' In the second, we show that \emph{controlled operations can be seen as charge-conserving operators} in a symmetry basis.

\emph{Charge-conserving operators as controlled operations:}
Let $\hat{Q}$ be a symmetric operator (i.e. charge-conserving). In its \emph{symmetry basis},\footnote{We defined a symmetry basis as the basis where an operator may become block diagonal, where each block is identified with a specific charge. The states in each charge sector are referred to as \emph{degenerate states} of the sector. The word ``degenerate'' is used in the sense that measurement of the states in a particular charge sector all give the same charge outcome, hence degenerate.} $\hat{Q}$ can be expressed as
\begin{equation}
\hat{Q} = \sum_{a,i_a, i'_a}  Q^{(a)}_{i_a, i'_a} \ket{i_a,a} \bra{i'_a,a},
\end{equation}
where $a$ labels the conserved charges and the indices $i_a, i'_a$ enumerate the \emph{degenerate states} of each charge $a$. This can be re-organized as
\begin{equation}
\hat{Q} = \sum_a \left(\sum_{i_a,i'_a} Q^{(a)}_{i_a,i'_a} \ket{i_a} \bra{i'_a}  \right) \otimes \ket{a}\bra{a}.
\end{equation}
The expression in the big brackets can be written succinctly as $\hat{Q}^{(a)}$: the operator acting in the degeneracy space of charge $a$. Therefore,
\begin{equation}
\label{Eq:SymmetricOpInSymmetryBasis}
\hat{Q} = \sum_{a} \hat{Q}^{(a)} \otimes \ket{a}\bra{a} .
\end{equation}
This is can be interpreted as a series of controlled operations, where the operator $\hat{Q}^{(a)}$---which has to be unitary---is applied if the charge is $a$. The circuit implementation will use two sets of qubits, one set called the (charge) control qubits to encode the charges and the other set called the (degeneracy) target qubits to encode the degenerate states. The $\ket{a}\bra{a}$ is a projector that acts on the (charge) control qubits, and it conserves charge. While $\hat{Q}^{(a)}$ is the corresponding unitary operator that acts on the (degeneracy) target qubits. The above equation says that if the (charge) control qubits are in a state with charge $a$, then apply operator $\hat{Q}^{(a)}$ on the (degeneracy) target qubits.

\emph{Controlled operations are charge-conserving operators:}
We now ``prove'' the reverse statement, that \emph{controlled operations can be represented as charge-conserving operators}.

This can be seen easily if the qubits are divided into two sets: the first set as target qubits and the second set as control qubits. The operations applied on the target qubits will depend on the charge of the control qubits. The operation on the control qubits can be written as projector $\ket{a} \bra{a}$ for each charge $a$---as the charge does not change, and the corresponding unitary operation on the target qubits as $\hat{Q}^{(a)}$, which depends on the charge $a$ of the control qubits. The whole operation can therefore be written as
\begin{equation}
    \hat{Q} = \sum_{a}  \hat{Q}^{(a)} \otimes \ket{a}\bra{a} ,
\end{equation}
which is exactly the form in Eq.~\ref{Eq:SymmetricOpInSymmetryBasis}. Therefore, controlled operations can indeed realize charge-conserving operators. 

The generic circuit for $\hat{Q}$ as written in Eq.~\ref{Eq:SymmetricOpInSymmetryBasis} is a series of controlled operations, where the specific controls applied depend on the charges. It is easy to check that $\hat{Q}$ can be rewritten as 
\begin{equation}
\label{Eq:Product form of Q}
\hat{Q} = \prod_a \left(\hat{Q}^{(a)} \otimes \ket{a}\bra{a} + \sum_{b \neq a} \mathbb{I} \otimes \ket{b}\bra{b} \right),
\end{equation}
where the expression in brackets means ``apply $\hat{Q}^{(a)}$ on target qubits if the control qubits encode charge $a$, otherwise do nothing.'' Then the product $\prod_a (\cdots)$ makes the circuit for $\hat{Q}$ a serial composition of controlled operations. 

Note that the results in this section are true in general for any symmetry group, and not only for $\mathbb{Z}_2$. The number of qubits needed for the encoding will depend on the number of conserved charges and their associated degeneracies. We do the example of two qubits below explicitly.

\subsubsection{Example: Two-qubit with $\mathbb{Z}_2$ symmetry.}
The two conserved charges on two qubits in $\mathbb{Z}_2$ basis are $\{0,1\}$, each with degeneracy two. So, this can be encoded easily with two qubits, with one encoding the charges and the other encoding the degenerate states. Therefore, a charge-conserving operator $Q$ on two qubits can be written, following the form in Eq.~\ref{Eq:SymmetricOpInSymmetryBasis}, as
\begin{equation}
\label{Eq:two-qubit symmetric op}
    \hat{Q} = \hat{Q}^{(0)} \otimes  \ket{0}\bra{0}  +  \hat{Q}^{(1)} \otimes \ket{1}\bra{1} ,
\end{equation}
where $\hat{Q}^{(0)}$ and $\hat{Q}^{(1)}$ are single-qubit unitaries associated to charges $0$ and $1$. Like in Eq.~\ref{Eq:Product form of Q}, 
\begin{align}
\hat{Q} &= \left( \hat{Q}^{(1)}\otimes  \ket{1}\bra{1} + \mathbb{I}  \otimes  \ket{0}\bra{0} \right)  \nonumber \\
&\qquad \qquad \left(\hat{Q}^{(0)} \otimes  \ket{0}\bra{0} + \mathbb{I} \otimes  \ket{1}\bra{1} \right),
\end{align}
which is realized as the circuit
\begin{equation}
\label{Eq:Symmetric operator circuit}
\begin{matrix}
    \Qcircuit @C=1em @R=.7em {
 & \multigate{1}{\hat{Q}} &  \qw \\
 & \ghost{\hat{Q}}        &   \qw  
}
\end{matrix}
=
\begin{matrix}
 \Qcircuit @C=1.5em @R=1.5em {
  & \gate{\hat{Q}^{(0)}} & \gate{\hat{Q}^{(1)}}  & \qw  \\
  & \ctrlo{-1}     & \ctrl{-1}       & \qw \\
}
\end{matrix}.
\end{equation}
In terms of matrix representation, we let $\hat{Q}^{(0)}$ and $\hat{Q}^{(1)}$ be single-qubit unitaries 
\begin{equation}
\label{Eq:Q1andQ2}
\hat{Q}^{(0)} = 
\begin{pmatrix}
a & c \\
b & d 
\end{pmatrix},
\quad 
\hat{Q}^{(1)} = 
\begin{pmatrix}
e & g \\
f & h 
\end{pmatrix},
\end{equation}
where the elements are not all independent. Then,
\begin{equation}
Q= 
    \begin{pmatrix}
    a & 0 & c & 0  \\
    0 & e & 0 & g  \\
    b & 0 & d & 0 \\
    0 & f & 0 & h
    \end{pmatrix}.
\end{equation}

This example can be generalized to more qubits. [The one-qubit case is trivial, $\hat{Q}^{(0)}$ and $\hat{Q}^{(1)}$ in Eq.~\eqref{Eq:two-qubit symmetric op} will be phases, so that $\hat{Q}$ is a $2\times2$ diagonal matrix.]

\subsection{$\mathbb{Z}_2$-symmetric basis transformations}
\label{Eq:Z2Transformation}
Since we specifically use $\mathbb{Z}_2$ to realize particle-conserving gates in this work, the gates that perform the basis transformation between the product basis and $\mathbb{Z}_2$-symmetric basis have to be obtained. For simplicity, we consider only neighbouring qubits. Extension to qubits that are not neighbouring will become obvious. 

To obtain this transformation, we apply the $\mathbb{Z}_2$ product (i.e. ``fusion'') rule on the computational basis states of two qubits $\{\ket{00}, \ket{01}, \ket{10}, \ket{11}\}$ is 
\begin{align}
0 \times 0 = 0; \quad 0 \times 1 = 1 \times 0 = 1; \quad 1 \times 1 = 0.
\end{align}
We see that there are two ways to obtain both charge $0$ and charge $1$, hence the degeneracy of both charge outcomes is two. Therefore, we write the $\mathbb{Z}_2$-invariant basis as $\{ \ket{0,0}, \ket{1,0}, \ket{0,1}, \ket{1,1} \}$, where the second entry is charge and the first enumerates the degenerate states associated with the charge.

To transform from the computational basis states to the $\mathbb{Z}_2$-invariant basis states, we use the map $F$: 
\begin{align}
& \ket{00} \rightarrow \ket{0,0},  \quad \ket{01} \rightarrow \ket{0,1},  \nonumber \\
& \ket{10} \rightarrow \ket{1,1}, \quad \ket{11} \rightarrow \ket{1,0} \nonumber.
\end{align}
(This is not the only possibility, but it gives the simplest gate possible.) If the map is ``read without the commas,'' it can be seen to be implementing a CNOT, where the first qubit is the control qubit and the second is the target qubit:
\begin{equation}
\label{Eq:FusionGate}
F = 
   \begin{matrix}
      \Qcircuit @C=1em @R=1em {
	& \ctrl{1} & \qw \\
       & \targ & \qw 
      }    
    \end{matrix} \quad .
\end{equation}
Although this is a common two-qubit gate, the interpretation it has in our usage is very important. On the left-hand of the gate the qubits are in a product basis, while on the right-hand side the qubits are now in a (charge-degeneracy) $\mathbb{Z}_2$ symmetry basis, $\{\ket{i_a, a}\}$ where the second qubit holds the charge and the first qubit the degenerate states of the charge.

\emph{Inverse basis transformation.} Since $F = \mathrm{CNOT}$ is an involution, the inverse basis transformation gate $S = F^{-1}$, also a CNOT gate.

\subsection{Generic particle-conserving gates}
\label{Sec:ApplyTwoSites}
In this section, we present circuits for generic particle-conserving gates, first for two qubits, and also for multiple qubits. We only consider single-excitation gates. 

\subsubsection{On two neighbouring qubits}
The circuit decomposition for a generic two-qubit gate is given in Refs.~\cite{vatan2004optimal, zhang2004minimum}. From that a particle-conserving gate and its circuit decomposition for two qubits can be obtained by applying the constraint of particle conservation. But rather than follow this approach, we shall use the procedure we provided above to construct the desired circuit for two-qubit particle-conserving operator, with the hope that it can also be applied to multiple qubits to yield efficient particle-conserving circuits.

We start with a $\mathbb{Z}_2$-symmetric gate, from which the particle-conserving gate can be obtained. Having derived all the required pieces, namely the gates for $F$, $Q$, and $S$ in sections Secs.~\ref{Sec:Symmetric Op quantum circuit} and \ref{Eq:Z2Transformation}, the generic circuit for an arbitrary $\mathbb{Z}_2$-symmetric unitary operator $Z$ on two qubits is derived by composing the gates, $Z=SQF$, to give 
\begin{equation}
\label{Eq:Generic Z2}
\begin{matrix}
    \Qcircuit @C=0.5em @R=.7em {
 & \multigate{1}{Z} &  \qw \\
 & \ghost{Z}        &   \qw  
}
\end{matrix} = 
   \begin{matrix}
        \Qcircuit @C=0.75em @R=1.5em {
	& \ctrl{1}    &  \gate{\hat{Q}^{(0)}}          &   \gate{\hat{Q}^{(1)}}    & \qw            & \ctrl{1} & \qw  \\
         & \targ        &  \ctrlo{-1}                   &  \ctrl{-1}           & \qw           &  \targ     & \qw  \\
        }        
    \end{matrix} \quad,
\end{equation}
for the $\hat{Q}^{(0)}$ and $\hat{Q}^{(1)}$ given in Eq.~\ref{Eq:Q1andQ2}. The matrix form of this circuit is 
\begin{equation}
Z = 
    \begin{pmatrix}
    a & 0 & 0 & c  \\
    0 & e & g  & 0  \\
    0 & f & h & 0 \\
    b & 0 & 0 & d
    \end{pmatrix}.
\end{equation}
Notice that the elements of $Q^{(1)}$ have moved to the center of $Z$ and those of $Q^{(0)}$ are pushed out to the corners.

The circuit for a particle-conserving operator can now be obtained from $Z$ by letting $b=c=0$, while the remaining variables are constrained to make the matrix unitary, namely that $a$ and $d$ will be purely phases, and the center block-matrix is unitary, where we let $f = -\exp({i\phi})g^*, h = \exp({i\phi})e^*$ for a phase $\phi$. The circuit above is not yet necessarily optimal in terms of the number of CNOTs and single-qubit gates for this generic case, but when we consider the specific parameterizations, the optimal circuit decomposition will be realized.


\subsubsection{On multiple qubits}
The idea used above for two qubits can be extended to multiple qubits. We consider the case of a $\mathbb{Z}_2$-symmetric and/or particle-conserving operator on two qubits that are not neighbouring, and without using swap gates. First, select the two qubits from a stack of quantum wires on which the symmetric operation is to be performed. Second, apply the CNOT transformation to bring the two qubits into a charge-degeneracy basis of $\mathbb{Z}_2$. Third, apply the controlled operations $C(\hat{Q}^{(0)})$ and/or $C(\hat{Q}^{(1)})$ on the two qubits. Finally, invert the initial basis transformation by applying the CNOT transformation again to transform back into the initial basis. An example circuit is shown 

\begin{equation}
\label{Eq:Long range excitation}
\begin{matrix}
    \Qcircuit @C=0.75em @R=1em {    
 & \sgate{Z^{(1,3)}}{3} &  \qw  \\
 & \qw       &   \qw  \\
  & \qw       &   \qw  \\
 & \gate{ Z^{(1,3)} }        &   \qw  
}
\end{matrix}
=
\begin{matrix}
        \Qcircuit @C=0.75em @R=1.5em {
            & \ctrl{3} &  \gate{\hat{Q}^{(0)}}   & \qw        & \gate{\hat{Q}^{(1)}}        & \ctrl{3}     & \qw  \\
	    & \qw     &  \qw                            & \qw        &     \qw                            & \qw          & \qw    \\
	    & \qw     &  \qw                            & \qw        &     \qw                            & \qw          & \qw    \\
            & \targ    &  \ctrlo{-3}                    & \qw        &    \ctrl{-3}                       &  \targ        & \qw       
        }      
\end{matrix},
\end{equation}
where it is understood that $Z$ is a function of some parameters through $\hat{Q}^{(0)}$ and $\hat{Q}^{(1)}$. The above circuit extends to any two qubits with any number of qubits in-between. This gate circuit can be used as part of the ``building blocks'' used in the construction of efficient brick-wall circuits with long-range qubit gates. We elaborate further on this later.

It is important to note that if the above circuit is adapted to particle conservation, it creates a single-excitation. To create higher excitations, for example say double excitations, the basis transformation idea we spelled out would need to be applied iteratively to bring all the participating qubits into the charge-degeneracy basis, then apply the charge-conserving operations, and finally undo all the basis transformation. But this is beyond the scope of our application to systems such as Heisenberg spin chains, which can be equivalently described by particle-conserving Hamiltonians with single-excitations.

\section{Types of particle-conserving gate parameterizations}
\label{Sec:PC gates}
We now provide some specific parameterizations of the generic particle-conserving  gate for any two qubits that may not necessarily be neighbouring. However, in this work we focus on only nearest-neighbouring (NN) and next-nearest neighbouring (NNN) qubits. Some of the gate types derived below have been use previously, others probably not.

\subsection{NN gate with two parameters in charge-1 block}
One example of a particle-conserving two-qubit gate on NN qubits is 
\begin{equation}
\label{gate A}
    A(\theta, \phi) = 
    \begin{pmatrix}
        1 & 0 & 0 & 0 \\
        0 & \sin{\theta} & e^{i\phi} \cos{\theta} & 0 \\
        0 & e^{-i\phi}\cos{\theta} & -\sin{\theta} & 0 \\
        0 & 0 & 0 & 1
    \end{pmatrix},
\end{equation}
which depends on two real parameters, rotation angles $\theta$ and $\phi$. Only the charge-1 block is parameterized. This can be obtained from the circuit Eq.~\ref{Eq:Generic Z2} by setting $\hat{Q}^{(0)}$ and $\hat{Q}^{(1)}$ as
\begin{equation}
\hat{Q}^{(0)} = \mathbb{I}, \quad  \hat{Q}^{(1)} = V(\theta, \phi),
\end{equation}
where
\begin{equation}
V(\theta, \phi) = 
\begin{pmatrix}
  \mathrm{sin} \theta & e^{i\phi} \mathrm{cos} \theta \\
e^{-i\phi} \mathrm{cos} \theta  & -\mathrm{sin} \theta\\
\end{pmatrix}.
\end{equation}
This can be decomposed as 
\begin{equation}
    V = XR_z(\phi) R_y(2\theta) R_z(\phi), 
    \label{Eq:DecompositionV}
\end{equation}
where $R_x(\theta) = e^{-i \frac{\theta}{2} X}$, $R_y(\theta) = e^{-i \frac{\theta}{2} Y}$, $R_z(\theta) = e^{-i \frac{\theta}{2} Z}$, and $X,Y,Z$ are the Pauli matrices $\sigma^x, \sigma^y, \sigma^z$. If we set $U = R_z(\phi) R_y(\theta)$, $V$ can be rewritten as
\begin{equation}
    V = U^{\dagger} X U.
\end{equation}
The complete circuit for $A(\theta, \phi)$ then is
\begin{equation}
\label{Eq:A gate}
\begin{matrix}
    \Qcircuit @C=0.5em @R=.7em {
 & \multigate{1}{A(\theta, \phi)  } &  \qw \\
 & \ghost{ A(\theta, \phi)  }        &   \qw  
}
\end{matrix} = 
 \begin{matrix}
        \Qcircuit @C=0.6em @R=1em {
	& \ctrl{1} & \gate{U(\theta, \phi)} & \targ     & \gate{U^{\dagger}(\theta, \phi)} & \qw & \ctrl{1} & \qw \\
           & \targ  &  \qw               &  \ctrl{-1} & \qw & \qw  & \targ & \qw \\
        }        
    \end{matrix}.
\end{equation}

\subsection{NN gate with two parameters in charge-$1$ and charge-$2$ blocks}
Another commonly used particle-conserving two-qubit gate on NN qubits is
\begin{equation}
\label{gate B}
    B(\theta, \phi) = 
    \begin{pmatrix}
        1 & 0 & 0 & 0 \\
        0 & \cos{\theta} & -i \sin{\theta} & 0 \\
        0 & -i \sin{\theta} & \cos{\theta} & 0 \\
        0 & 0 & 0 & e^{i \phi}
    \end{pmatrix},
\end{equation}
which depends on two real parameters, say $\theta$ in charge-$1$ block and $\phi$ in charge-$2$ block. This can be obtained from the circuit Eq.~\ref{Eq:Generic Z2} if we set 
\begin{equation}
\hat{Q}^{(0)} = P(\phi), \quad  \hat{Q}^{(1)} = R_x(2\theta),
\end{equation}
where
\begin{equation}
\label{Eq:Phase and Rx gates}
P(\phi) = 
\begin{pmatrix}
    1 & 0 \\
    0 & e^{i\phi}
\end{pmatrix},
\quad 
R_x(\theta) = e^{-i \frac{\theta}{2} X}.
\end{equation}
The complete circuit for $B(\theta, \phi)$ from Eq.~\ref{Eq:Generic Z2} is
\begin{equation}
\begin{matrix}
    \Qcircuit @C=0.5em @R=.7em {
 & \multigate{1}{B(\theta, \phi)} &  \qw \\
 & \ghost{B(\theta,\phi)}        &   \qw  
}
\end{matrix} = 
   \begin{matrix}
        \Qcircuit @C=0.5em @R=1.5em {
	   & \ctrl{1}    &  \gate{P(\phi)}          &   \gate{R_x(2\theta)}   & \qw            & \ctrl{1} & \qw  \\
            & \targ        &  \ctrlo{-1}                   &  \ctrl{-1}           & \qw           &  \targ     & \qw   \\
        }        
    \end{matrix} \quad.
\end{equation}
The circuit can be made more efficient, as the number of CNOTs can be reduced at the cost of introducing additional single-qubit operations. The circuit is equivalent to this
\begin{equation} 
   \begin{matrix}
        \Qcircuit @C=0.5em @R=1.5em {
	   & \gate{P(\phi/2)} & \ctrl{1}    &  \qw         &   \gate{R_x(2\theta)}   & \qw            & \ctrl{1} & \qw  \\
            & \gate{P(\phi/2)} & \targ        & \gate{P(-\phi/2)}                   &  \ctrl{-1}           & \qw           &  \targ     & \qw   \\
        }        
    \end{matrix} \quad.
\end{equation}
This can be seen not to introduce any phase factor except when both qubits are set, where it gives the phase $e^{i\phi}$.

\subsection{NN two-qubit general particle-conserving gate}
\label{Sec:Generalized PC gate}
We present the general particle-conserving gate $G(\alpha, \theta, \phi_1, \phi_2) $ on NN qubits as
\begin{equation}
\label{gate B}
    G = 
    \begin{pmatrix}
        1 	& 	0  					& 	0 	&	 0 \\
        0 	& 	e^{i\alpha} e^{i\frac{\phi_1+\phi_2}{2}}\cos{\theta} 	& 	e^{i\alpha} e^{i\frac{\phi_1 - \phi_2}{2}}\sin{\theta} 	&	 0 \\
        0 	& 	-e^{i\alpha} e^{-i\frac{\phi_1 - \phi_2}{2}} \sin{\theta} 	&  e^{i\alpha} e^{-i\frac{\phi_1+\phi_2}{2}} \cos{\theta} 	& 0 \\
        0 	& 	0 					& 	0 	& 						1
    \end{pmatrix},
\end{equation}
which depends on only $4$ real parameters $\alpha, \theta, \phi_1, \phi_2$, which are all in the charge-$1$ block. The proof that this gate is the most general of the particle-conserving gates on NN two qubits is shown in Appendix ~\ref{Sec:ProofofGeneralPCgate}. It is clear that the gates $A$ and $B$ given above are instances of $G$.

If we set the controlled operators of Eq.~\ref{Eq:Generic Z2} as $\hat{Q}^{(0)} = \mathbb{I}$ and  $\hat{Q}^{(1)}$ as the 1-qubit general unitary
\begin{equation}
U(\alpha, \theta, \phi_1, \phi_2) = e^{i \alpha} R_z(\phi_1) R_y(2 \theta) R_z(\phi_2),
\end{equation}
then the circuit implementing $G(\alpha, \theta, \phi_1, \phi_2)$  becomes 
\begin{equation}
\begin{matrix}
    \Qcircuit @C=0.5em @R=.7em {
 & \multigate{1}{G(\alpha, \theta, \phi_1, \phi_2)} &  \qw \\
 & \ghost{G(\alpha, \theta, \phi_1, \phi_2)}        &   \qw  
}
\end{matrix} = 
   \begin{matrix}
        \Qcircuit @C=1em @R=1.5em {
            & \ctrl{1}       & \qw      &   	 \gate{U}        & \qw            & \ctrl{1}  & \qw \\
            & \targ      & \qw       & 	 \ctrl{-1}           & \qw           &  \targ      & \qw  \\
        }        
    \end{matrix}.
\end{equation}
Since $U$ can be written as $U(\alpha, \theta, \phi_1, \phi_2)  = e^{i \alpha} AXBXC$, where
\begin{align}
&A = R_z(\phi_1) R_y(\theta) \nonumber \\
&B = R_y(-\theta) R_z\left(-\frac{\phi_1 + \phi_2}{2}\right) \\
&C = R_z\left(\frac{\phi_2 - \phi_1}{2}\right) \nonumber,
\end{align} 
and such that $ABC = \mathbb{I}$, the controlled-U operation can be decomposed as 
\begin{equation}
   \begin{matrix}
        \Qcircuit @C=1em @R=1.5em {
            &   \gate{U}  &    \qw  \\
            &  \ctrl{-1}      &  \qw 
            }
   \end{matrix} = 
     \begin{matrix}
        \Qcircuit @C=0.5em @R=1.5em {
            & \gate{C} & \targ  &  \gate{B}  &  \targ          &  \gate{A}            & \qw & \qw  \\
            & \qw & \ctrl{-1}  & \qw          &  \ctrl{-1}        &  \qw           &  \gate{P(\alpha)}     & \qw  
        }        
    \end{matrix},
\end{equation}
with the single qubit gates $A, B, C$ given above, and $P(\alpha)$ is a phase gate like in Eq.~\ref{Eq:Phase and Rx gates}.

Although gate $G$ is ``costlier'' than $A$ and $B$ in terms of the number of parameters and number of CNOTS, but there might be situations where using $G$ can provide some advantages over using $A$ or $B$, for example, where there is demand for higher optimization accuracy.

\subsection{Long-range single excitation two-qubit gates}
The parameterizations shown above can be extended to cases where the two qubits may not be nearest neighbours, hence allowing for the possibility of creating long-range single excitations on two distant qubits. We do this for the example of three qubits using the same parameterization as for the A-gate above.

Let $A^{(1,3)}$ be the gate which creates a NNN single-excitation on three qubits (denoted respectively with the indices $(1,3)$). Of the basis states of three qubits, the ones with the desired single excitations are $\{\ket{001}, \ket{100}\}$ with only one particle, and  $\{\ket{011}, \ket{110}\}$ with two particles. In both sets, a particle is either created on the first qubit and annihilated on the third qubit, or vice versa, while the state of the middle qubit is unchanged, which means nothing happens to the second qubit. The two-qubit case can therefore be extended to obtain the gate $A^{(1,3)}$ as
\begin{equation}
\label{Eq:Long range excitation}
\begin{matrix}
    \Qcircuit @C=0.75em @R=1em {    
 & \sgate{A^{(1,3)}}{2} &  \qw  \\
 & \qw       &   \qw  \\
 & \gate{ A^{(1,3)} }        &   \qw  
}
\end{matrix}
=
\begin{matrix}
        \Qcircuit @C=0.75em @R=1.5em {
            & \ctrl{2}  &  \gate{U^{\dagger}(\theta, \phi)} & \targ     & \gate{U(\theta, \phi)}  & \ctrl{2}  & \qw  \\
	    & \qw  	   &\qw        &     \qw               & \qw           & \qw         &\qw    \\
            & \targ      & \qw       & 	 \ctrl{-2}           & \qw           &  \targ      & \qw       
        }      
\end{matrix},
\end{equation}
where the gate $A^{(1,3)}$ is understood to be a function of $\theta$ and $\phi$. It is easy to check that the circuit decomposition will generate the right (and desired) transformation on the basis states of three qubits, with non-trivial rotations only between the basis states participating in the excitation. This gate construction makes use of only three CNOTS and two single-qubit gates.

It should be noted that no swap gates have been employed in creating the gate in Eq.~\ref{Eq:Long range excitation}. There is of course an alternative na\"ive approach that could have been used to create the gate: first, apply a swap gate on the first and second qubits, then apply the two-qubit gate $A(\theta, \phi)$ on the second and third qubits, and finally apply another swap gate again on the first and second qubits to restore back the qubits ordering. This gate achieves the same transformation as the $A^{(1,3)}$ above, but it requires nine CNOTs and two single-qubit operations. Therefore, the gate in Eq.~\ref{Eq:Long range excitation} is more efficient.

Finally, it is clear how to extend this example to exploit the other parameterizations, and even to consider more than $3$ qubits. In fact, the gate parameterizations can be applied to boundary qubits to impose periodic boundary condition.

\section{Particle-Conserving Brick-wall Circuits}
\label{Sec:PQCs}
In this  section, we construct particle-conserving brick-wall parameterized quantum circuits (PQCs). First, we will recall how quantum states that conserve particle number can be defined. Secondly, we use the particle-conserving gates in the previous section to construct parameterized quantum circuits that conserve particle number. These parameterized circuits are used as ansatzes for particle-conserving quantum states, which through variational optimization can yield the ground states of particle-conserving Hamiltonians. In the next section (Sec.~\ref{Sec:Test Model}), we investigate the performance of the circuits with Heisenberg spin models with nearest and next-nearest interactions.

\subsection{Particle-conserving quantum states}
A particle-conserving quantum state is mathematically defined in a Fock space. For a system of $N$ spinless electrons on a lattice of $L$ sites, the Fock space $\mathcal{H}_{N,L}$  is spanned by the occupation number basis states $\{ \ket{n_1, \ldots, n_L}\}$,
where $n_i \in \{ 0, 1 \}$ such that $\sum_{i=1}^L n_i = N$. The dimension of $\mathcal{H}_{L,N}$ is $d_{N,L} = \left( \begin{matrix} L \\ N \end{matrix} \right)$.

A general pure quantum state can be written in the occupation number basis as 
\begin{equation}
\ket{\Psi_{N,L}} = \sum_{n_1,  \ldots, n_L} c_{n_1,  \ldots, n_L} \ket{n_1,  \ldots, n_L}, 
\end{equation}
with $d_{N,L}$ number of complex coefficients $c_{n_1, \ldots, n_L} \in \mathbb{C}$. By applying the constraint of normalization and neglecting a global factor, the number of coefficients can be reduced to $2(d_{N,L} -1) $ real numbers.

We assume that an ansatz for $\ket{\Psi_{N,L}}$ can be created on a quantum processor by applying a sequence of local gates on an initial state $\ket{\psi_0}$. For hardware-efficient ansatzes, we let the local gates be one- and two-qubit gates on nearest-neighbouring (NN) and next-nearest-neighbouring (NNN) qubits. The number of gates that will be used in the circuit construction is such that the total number of parameters in the circuit is adequate to span the Fock space.

\subsection{Brick-wall particle-conserving circuits}
\label{Sec:U1 symmetric quantum circuit}
We now consider the construction of particle-conserving parameterized quantum circuits. One way to construct a quantum circuit that conserves particle number is with gates that conserve particle number locally. To this end, we use the particle-conserving gate types derived in Sec.~\ref{Sec:PC gates} as ``building blocks'' to construct a family of brick-wall PQCs. The circuits are homogeneous in nature, involving tiles of gates of only parameterization type. The circuits are named after the gate types used to build them. For example, the circuit built with gate type A is named $C_A$.

Let $\ket{\Psi(\pmb{\theta})}$ be a parameterized ansatz state created on a quantum processor by the application of a particle-conserving unitary gate $\hat{U}(\pmb{\theta})$ on a reference state $\ket{\psi_0}$,
\begin{equation}
\ket{\Psi(\pmb{\theta})} = \hat{U}(\pmb{\theta}) \ket{\psi_0},
\end{equation}
where $\pmb{\theta}$ is the vector of all parameters of the ansatz state. Our reference state $\ket{\psi_0}$ will be initialized by setting $N$ qubits to the state $\ket{1}$ and the remaining $(L-N)$ qubits to the state $\ket{0}$. Putting a qubit into state $\ket{1}$ is simply by the application of an X gate.

 An hardware-efficient ansatz for the unitary gate $\hat{U}(\pmb{\theta})$ will, in general, be a composition of local gates, which optimize the use of hardware resources with respect to the number of qubits and device topology.  An important question to consider to create a parameterized circuit is to determine the number of local gates needed to ``tile'' the circuit. We answer this question for gates that introduce only two parameters to a circuit, like gates $A(\theta, \phi)$ or $B(\theta, \phi)$ in Sec.~\ref{Sec:PC gates}. Loosely speaking, the number of such gates is that which is sufficient to span the Fock space. For brick-wall circuits, the gates will be arranged into alternate layers.
 
In the following, we present algorithms to construct particle-conserving parameterized brick-wall circuits with NN and NNN gates.

\subsubsection{With only NN gates}
We first consider two-parameter NN two-qubit gates [such as $A(\theta, \phi)$ or $B(\theta, \phi)$] to build a circuit for $\hat{U}(\pmb{\theta}) $ in a brick-wall alternating pattern. Since each of these gates introduces only two real parameters, then $(d_{N,L} - 1)$ number of basic gates should in principle be sufficient. This would give a total of $2(d_{N,L} -1)$---the minimum number of real parameters needed.  However, it was found in Ref.~\cite{gard2020efficient} that using this number of basic gates does not give a quantum circuit that spans the N-particle Fock space. Rather they found numerically that a total of $d_{N,L}$ number of basic gates are needed, which will bring the total number of parameters to $2d_{N,L}$---two more than necessary. The number of parameters can then be reduced by fixing two of them. Notwithstanding, we discovered that $(d_{N,L}-1)$ number of gates ($A$ or $B$) is indeed sufficient to span the space, albeit at the expense of introducing two additional swap operations, and hence less efficient. Therefore, we take $d_{N,L}$ as the optimal number of gates needed. 

\begin{figure}
\[ \Qcircuit @C=1em @R=1em {
\ket{0} &	&	\qw	&	\multigate{1}{M}	&	\qw	&	\qw	&		&		&		&	\qw	&	\multigate{1}{M}	&	\qw	&	\qw	\\
\ket{0} &	&	\gate{X}	&	\ghost{M}	&	\multigate{1}{M}	&	\qw	&		&		&		&	\qw	&	\ghost{M}	&	\multigate{1}{M}	&	\qw	\\
\ket{0} &	&	\qw	&	\qw	&	\ghost{M}	&	\qw	&		&		&		&	\qw	&	\qw	&	\ghost{M}	&	\qw	\\
	&		&		&		&		&		&		&		&		&		&		&		\\
\vdots	&		&		&	\vdots	&		&		&	\ddots	&		&		&		&		&	\vdots	\\
	&		&		&		&		&		&		&		&		&		&		&		\\
\ket{0} &	&	\gate{X}	&	\qw	&	\multigate{1}{M}	&	\qw	&		&		&		&	\qw	&	\qw	&	\multigate{1}{M}	&	\qw	\\
\ket{0} &	&	\qw	&	\multigate{1}{M}	&	\ghost{M}	&	\qw	&		&		&		&	\qw	&	\multigate{1}{M}	&	\ghost{M}	&	\qw	\\
\ket{0} &	&	\gate{X}	&	\ghost{M}	&	\qw	&	\qw	&		&		&		&	\qw	&	\ghost{M}	&	\qw	&	\qw
} \]
\caption{A brick-wall parameterized quantum circuit with ``placeholder'' gates $M$. We choose M from the list of gates $\{ A,B,G\}$ presented in Sec.~\ref{Sec:PC gates}. The initial $N$ number of X gates is to bring the quantum processor into the $\mathcal{H}_{N,L}$ Fock space (as explained in the text), and then followed by an alternating sequence of odd and even gates. In general, the gates $M$ all have different parameters.}
\label{Fig:QuantumCircuit_A}
\end{figure}
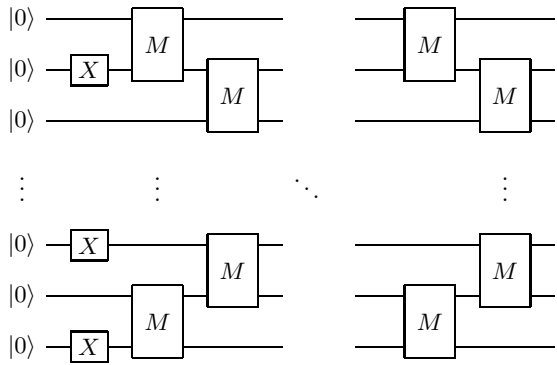

We now present the brick-wall particle-conserving parameterized quantum circuits. Let the circuit be denoted $C_M (\pmb{\theta})$, constructed by composing particle-conserving gates together. We choose $M$ from the set of gates $\{ A, B, G\}$, a set of NN two-qubit gates. The schematic diagram for the circuit $C_M $ is shown in Fig.~\ref{Fig:QuantumCircuit_A}. For simplicity, we stick with using $d_{N,L}$ number of basic gates, with all free parameters. The algorithm behind the construction of this circuit is presented:
\begin{itemize}
\item Apply X gates to N number of qubits to bring the quantum processor into the N-particle subspace.
\item Apply a first half-layer of M gates (viz. $A$, $B$, or $G$) on odd sites, followed by another half-layer on even sites, making up the first layer. 
\item Repeat the second step until the minimum number of layers, $\# l$, required to create the desired circuit is achieved. We determine this number from $\# l = \lceil \frac{d_{N,L}}{L-1} \rceil$.\footnote{The idea behind the formula is that, since only $(L-1)$ number of NN two-qubit gates ``tiles'' a layer, then to have a circuit with $d_{N,L}$ number of gates, a total of $d_{N,L}/(L-1)$ layers will be needed. To avoid fractions, we round up. But this will increase the number of gates needed, and therefore the number of parameters. The extra parameters can be fixed to bring down the parameter count.}
\end{itemize}

\subsubsection{Both NN and NNN gates}
\label{Sec:NN and NNN circuits}
We now extend the algorithm above to the case where the circuit does not only consists of NN gates but also NNN gates. Again, we do it layer-by-layer. But now the layers are those with NN and NNN gates in alternate patterns. For example, if the first layer consists of NN gates, the second layer will consist of NNN gates, and so on. If the parameters are made small, the circuit can be seen as some Trotter decomposition of the exponential of some underlying Hamiltonian with NN and NNN interactions.

\begin{figure}
\Qcircuit @C=1em @R=1em {
\ket{0} &  & \qw  & \multigate{1}{M} & \qw & \sgate{M}{2}  & \qw  & \qw & \multigate{1}{M} & \qw  & \qw \\
\ket{0} &  & \gate{X}  & \ghost{M} & \multigate{1}{M} & \qw  & \sgate{M}{2}  & \qw & \ghost{M}  & \multigate{1}{M} & \qw \\
\ket{0} &  & \qw & \multigate{1}{M} & \ghost{M}  &  \gate{M} & \qw & \sgate{M}{2} & \multigate{1}{M} & \ghost{M} & \qw \\
\ket{0} &  & \qw & \ghost{M}  &  \multigate{1}{M} &  \sgate{M}{2} & \gate{M} & \qw & \ghost{M} & \multigate{1}{M} & \qw \\
\ket{0} &  & \gate{X} & \multigate{1}{M}  &  \ghost{M}  & \qw & \sgate{M}{2}  & \gate{M} & \multigate{1}{M} & \ghost{M} & \qw \\
\ket{0} &  & \qw & \ghost{M}  &  \multigate{1}{M} &  \gate{M} & \qw  & \sgate{M}{2} & \ghost{M} & \multigate{1}{M} & \qw \\
\ket{0}  & & \qw  & \multigate{1}{M}  &  \ghost{M}  & \qw  &\gate{M}  & \qw & \multigate{1}{M} & \ghost{M} & \qw \\
\ket{0} &  & \gate{X}  & \ghost{M} &  \qw & \qw & \qw  & \gate{M} & \ghost{M} & \qw & \qw
} 
\caption{Parameterized quantum circuit for $3$ particles on $8$ sites with both NN and NNN gates. This creates an ansatz state in the Fock space $\mathcal{H}_{3,8}$. The circuit consists of three alternating layers of NN and NNN gates. The first layer consists of NN gates, the second layer consists of NNN gates, and the third layer consists of another NN gates. Note that the gates $M$s are functions of some parameters. To reduce computational cost, we will choose $M$ as either $A(\theta, \phi)$ or $B(\theta, \phi)$.}
\label{Fig:Circuit Example with NN and NNN gates}
\end{figure}
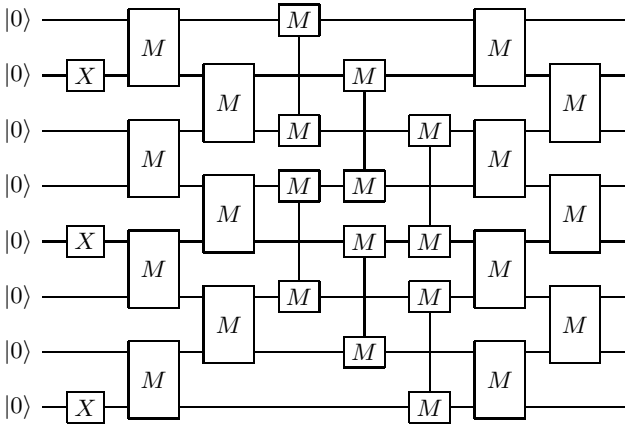

For NN qubits, the gates were placed alternately on even and odd sites. Even and odd sites are those whose site indices are equivalent to $0~(\mathrm{mod} ~2)$ and $1~(\mathrm{mod} ~2)$ respectively. Therefore, the placement of gates with NNN qubits would be determined $\mathrm{mod}~3$. This implies a NNN layer would consists of three sub-layers. The algorithm for a single NNN layer is
\begin{itemize}
\item  Apply a first $1/3$-layer of gates [e.g. $A^{(1,3)}(\theta, \phi)$] on sites whose indices are $0$ ($\mathrm{mod} 3$).
\item Follow by a second $1/3$-layer of gates applied to sites whose indices are $1$ ($\mathrm{mod} 3$).
\item Finally, end with a third $1/3$-layer of gates applied to sites whose indices are $2$ ($\mathrm{mod} 3$).
\end{itemize}
If we combine this sub-algorithm with the previous one, we would construct a brick-wall circuit with alternate layers of NN and NNN gates. We repeat the algorithm until the desired number of layers is achieved. An example of an $8$-qubit brick-wall circuit with NN and NNN gates is given in Fig.~\ref{Fig:Circuit Example with NN and NNN gates}. This example circuit creates an ansatz state in the Fock space $\mathcal{H}_{3,8}$. The circuit consists of three layers, with the first layer consisting of NN gates, the second layer consists of NNN gates, and the third layer consists of another NN gates. This circuit with both NN and non-NN gates can be compared to one with only NN gates, where the second layer would be made up of NN gates. Since the circuit in example Fig.~\ref{Fig:Circuit Example with NN and NNN gates} consists of only three layers, this by no means spans the Fock space $\mathcal{H}_{3,8}$. This example can be generalized to other cases by either varying the number of sites and/or particle number. We will refer to this class of circuits with the name $C_M^{\mathrm{ex}}$, where the ``ex'' is used to refer to ``extended,'' i.e. beyond nearest-neighbouring, and $M$ is the gate label which can be any particle-conserving gate parameterization, for example $\{A, B, G\}$.

\section{Numerical experiments with particle-conserving circuits}
\label{Sec:Test Model}
In this section, we present results of numerical experiments performed with particle-conserving brick-wall circuits for two different learning problems: ground state of spin chain Hamiltonians with and without next-nearest-neighbouring interactions, and learning of random quantum states. The gates making up the particle-conserving circuits are chosen from the parameterizations given in Sec.~\ref{Sec:PC gates}. The aim of the experiments is to test the learning capacity of the gates, and the relative advantage of using circuits with long-range gates over their counterparts with only nearest-neighbouring gates.

\subsection{Cost functions and averaging procedure}
Before presenting the results, we state the two cost functions that we use in this work, namely energy and fidelity. 

\subsubsection{Minimization of Energy}
The ground state of a Hamiltonian $H$ can be obtained from a variational algorithm by minimizing energy
\begin{equation}
    E(\pmb{\theta^*}) = \operatorname*{argmax}_{\pmb{\theta}}  \bra{\psi (\pmb{\theta})} H \ket{\psi (\pmb{\theta})},
\end{equation}
with respect to the parameters $\pmb{\theta}$ to obtain the minimum point $\pmb{\theta^*}$. The experiment is repeated $N_T$ times. The final ground state energy is average of the trials
\begin{equation}
\bar{E} = \substack{\mathbb{E} \\ {\pmb{\theta^*} }} ~ E(\pmb{\theta^*}).
\end{equation}

\subsubsection{Maximization of Fidelity}
We also use the circuits to learn statevectors that are randomly sampled from some $N$-particle Fock spaces. The learning is done independently for each sample, and the results averaged.

The learning can be quantified through fidelity for a pure state as
\begin{equation}
F_i(\pmb{\theta}) =  | \langle \phi_i \ket{\Psi_i (\pmb{\theta})} |^2, \quad i = 1,2,\ldots, N_S
\end{equation}
where $\ket{\Psi_i (\pmb{\theta})}$ is the circuit ansatz state and $\ket{\phi_i}$ is the  sample state, and $N_S$ is the number of samples. The optimization task for a single sample $i$ is
\begin{equation}
F_i(\pmb{\theta^*}) = \operatorname*{argmax}_{\pmb{\theta}} F_i(\pmb{\theta}),
\end{equation} 
where $\pmb{\theta^*}$ is the optimized parameter point for a single trial, and $F_i(\pmb{\theta^*})$ is the optimized fidelity value, whose theoretical value is $1$. Let the (uniform) average over the $N_T$ trials for a single sample be  
\begin{equation}
\bar{F}_i = \substack{\mathbb{E} \\ {\pmb{\theta^*} }} ~ F_i(\pmb{\theta^*}).
\end{equation}
Furthermore, let the average fidelity for all $N_S$ samples, i.e. fidelity per sample per trial, be 
\begin{equation}
    \bar{F} = \frac{1}{N_S} \sum_{i = 1 }^{N_S} \bar{F}_i .
\end{equation}
To quantify the errors in optimization, we define (i) relative error per sample  $\epsilon_i = 1 - \bar{F}_i$, and (ii) relative error per sample per trial $\bar{\epsilon} = \frac{1}{N_S} \sum_i \epsilon_i$.

The last important details for the simulations are: For each set of experiments, the circuits are initialized from the same point in the parameter space. The initial values of the parameters are chosen independently and identically from the uniform distribution $\mathcal{U}[-\pi, \pi)$. We used COBYLA (classical) optimizer for the classical optimization, where the number of optimization steps is set according to the problem size. The numerical experiments were done on a noiseless quantum simulator. The codes were written with (IBM) Qiskit.\cite{Qiskit}

\subsection{Ground state of Heisenberg XXZ model}
We first investigate the Heisenberg XXZ model with the parameterized quantum circuits with only NN gates. The Hamiltonian is
\begin{equation}
\label{Eq:XXZModel}
    H = \sum_{i=1}^{L-1} \left( X_i X_{i+1} + Y_i Y_{i+1} + \gamma Z_i Z_{i+1} \right),
\end{equation}
where $X,Y,$ and $Z$ are the usual Pauli matrices $\sigma^x, \sigma^y,$ and $\sigma^z$. This model conserves total magnetization, $M = \sum_i Z_i$, and therefore $[H,M]=0$. Equivalently, the model can be viewed to conserve particle number due to the mapping between the XXZ model and hardcore Bose-Hubbard model (HCBH). This is reviewed in Appendix \ref{App:XXZ2BHM}. Therefore, this Hamiltonian can be simulated with a particle-conserving quantum circuit.

The simulations of the Hamiltonian (with $\gamma = 1$) are performed with brick-wall circuits $C_A$, $C_B$, and $C_G$ with only NN gates. We consider the model on only $L = 4,6,8$ sites, and also its XX reduction (when $\gamma=0$) on $L=8$ sites. The estimate of the expectation value was computed with $n_\mathrm{shots}=1024$ shots. We set the number of function evaluations in the optimizer to be such that the ratio of the number of evaluations to the Fock space dimension should scale roughly the same for all the lattice sizes considered. For $L=4$ the number is set equals to $1000$.

\begin{figure}
    \centering
    \subfloat[$L = 4$ sites]{\includegraphics[width=0.5\columnwidth]{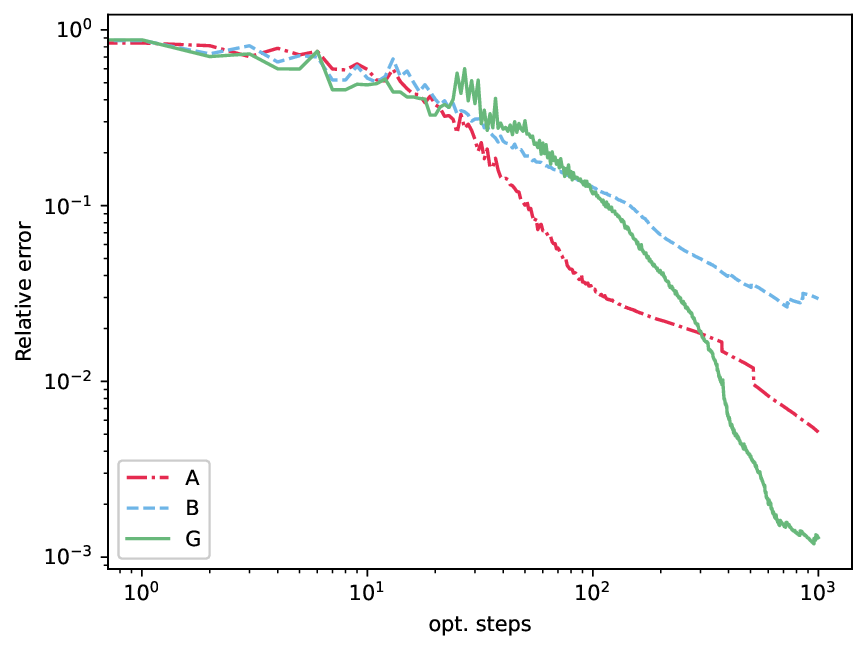}} 
    \hfill
    \subfloat[$L = 6$ sites]{\includegraphics[width=0.5\columnwidth]{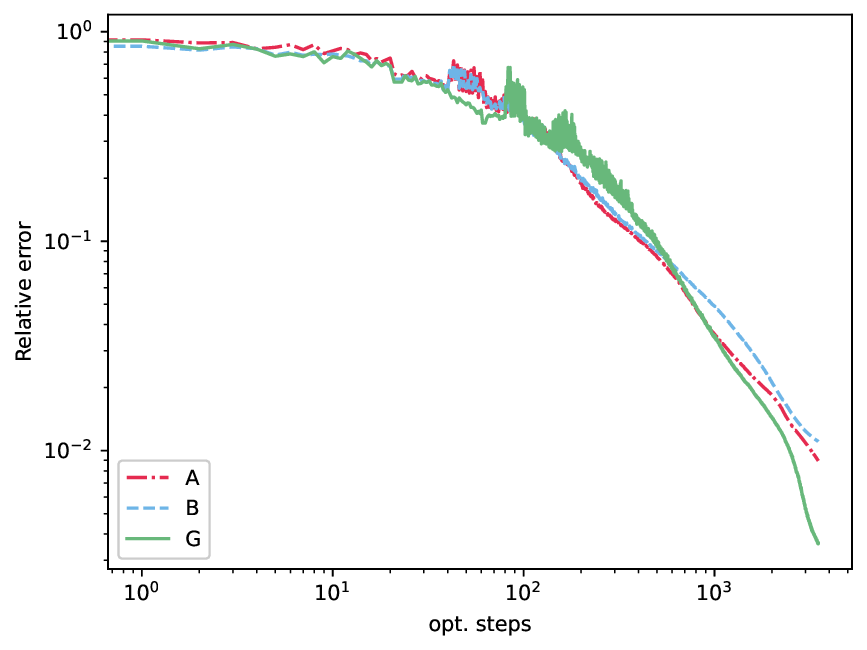}}
    
     \subfloat[$L = 8$ sites]{\includegraphics[width=0.5\columnwidth]{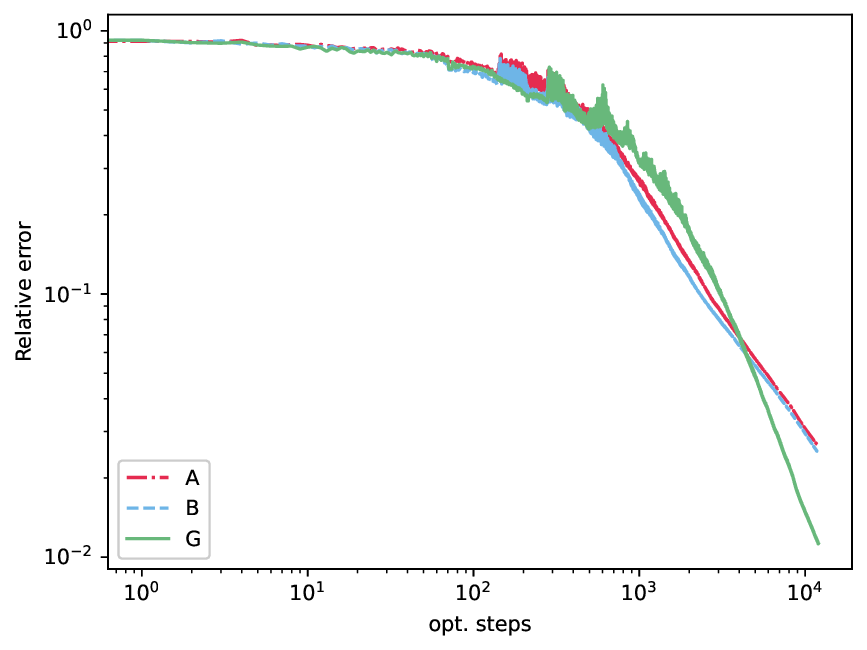}}
     \hfill
    \subfloat[$L = 8$ sites, XY model]{\includegraphics[width=0.5\columnwidth]{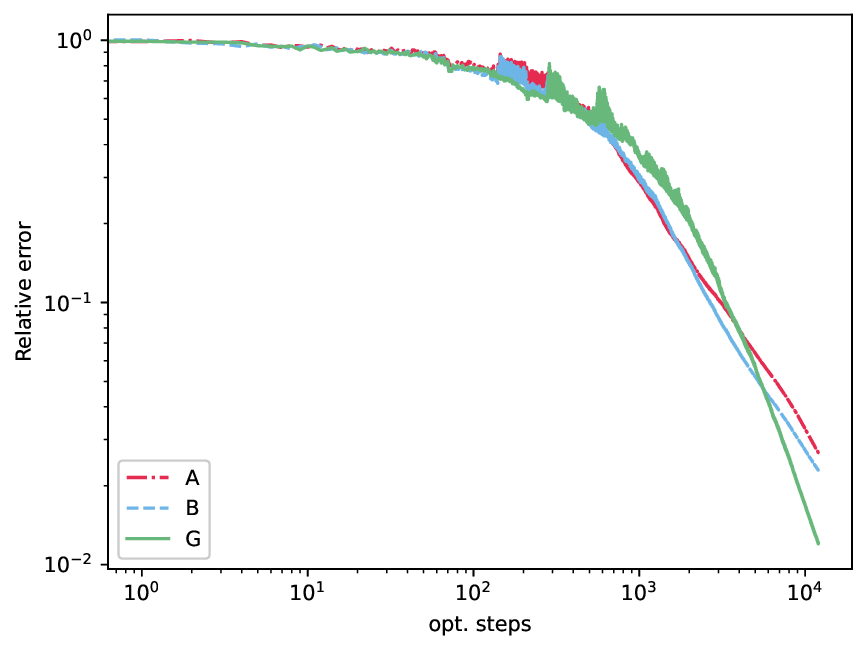}}     
\caption{Results of the variational optimization of Heisenberg model $(a)$-$(c)$ for lattice sizes $L=4,6,8$ and its XX reduction (d) for $L=8$ sites. The horizontal axis (``opt. steps'') is the number of classical optimization steps (or function evaluations), and the vertical axis (``relative error'') is the error relative to the true ground state energy. In all the plots, the red ``dashed-dotted''  line is for the A-gate, while the blue dashed line is for the B-gate, and the green solid line is for the G-gate.}  
\label{Fig:LearningHeisenbergModel}
\end{figure}

The results of the error in learning the ground states are presented in Fig.~\ref{Fig:LearningHeisenbergModel}. While it can be seen that circuit $C_A$ performs better than circuit $C_B$ for most cases, circuit $C_G$ outperforms both in all cases. The lowest values obtained for the sizes considered are collected in Table~\ref{Table:Energies}. As evident from the table, none of the circuits achieved the global minimum in all cases, even for circuit $C_G$, which is made of the generalized particle-conserving gate $G$. This can be attributed to two reasons: $(1)$ The limitation of the expressiveness of the brick-wall circuit structure. $(2)$ The effect of barren plateau phenomenon, which beside other causes, is also known to affect circuits initialized from a uniform distribution.\cite{wang2023trainability} 

\begin{table}
\begin{tabularx}{\columnwidth} { 
 | >{\centering\arraybackslash}X 
 | >{\centering\arraybackslash}X 
 | >{\centering\arraybackslash}X 
  | >{\centering\arraybackslash}X 
 | >{\centering\arraybackslash}X | }
 \hline
  &  \multicolumn{3}{|c|}{ XXZ } & XX \\
 \hline
 $L$  &  $4$ & 6 & 8 & 8  \\
 \hline  
$E_0$  &   $-6.4641$  & $-9.9743 $  & $-13.4997$ & -9.5175  \\
\hline
$E_A$   & $-6.4308$  & $-9.8851$ & $-13.1431$ & -9.2626 \\
$ E_B$  & $-6.2734$  & $-9.8640$ & $-13.1648$  &  -9.2991\\
$ E_G$  & $-6.4558$  & $-9.9383$ & $-13.3480$   &  -9.4032\\
 \hline 
\end{tabularx}
\caption{Table shows a comparison between the true ground state energy, $E_0$, of Heisenberg models (XXZ and XX models) versus average values obtained from numerics for lattice sizes $L=4,6,8$. The obtained average energy for the different particle-conserving circuits $C_A$, $C_B$, and $C_G$ are $E_A$, $E_B$, and $E_G$ respectively.}
\label{Table:Energies}
\end{table}

\subsection{Heisenberg model with NN and NNN interations}
Here, we compare the efficiency of circuits with and without long-range gates, that is, between a circuit with only NN two-qubit gates and another with both NN and NNN two-qubit gates. We simulate the Heisenberg spin chain with next-nearest interactions. The Hamiltonian can be written as
\begin{equation}
H = \sum_{i=1}^{L-1} \sigma_i \sigma_{i+1}  + \sum_{i=1}^{L-2} \sigma_i \sigma_{i+2},
\end{equation}
where $L$ is the number of lattice sites, with open boundary condition, and $\sigma_i = (X_i, Y_i, Z_i)$. The Hamiltonian conserves particle number, and can therefore be investigated with particle-conserving circuits. We compare the following two circuits: $(i)$ $C_A$, consisting of only NN gates, and $(ii)$ $C_A^{\mathrm{ex}}$, consisting of both NN and NNN gates (as discussed in Sec.~\ref{Sec:NN and NNN circuits}). For fair comparisons, the two circuits are subjected to the same constraints: same number of layers, same number of parameters,\footnote{Since circuits $C_A$ have more gates than $C_A^{\mathrm{ex}}$, the former circuit will \emph{a priori} have more parameters than the latter. To make the number of parameters in the two circuits the same, we adopt a convention that the last extra parameters in $C_A$ are fixed all equal.} and hence both circuits are initialized from the same point in parameter space in each experiment. We limit our investigation to circuits with only ``type-A'' parameterization (see Sec.~\ref{Sec:PC gates}) and for $L=8$ qubits. 

\begin{figure}
    \centering
    \subfloat[$m = 3$]{\includegraphics[width=0.5\columnwidth]{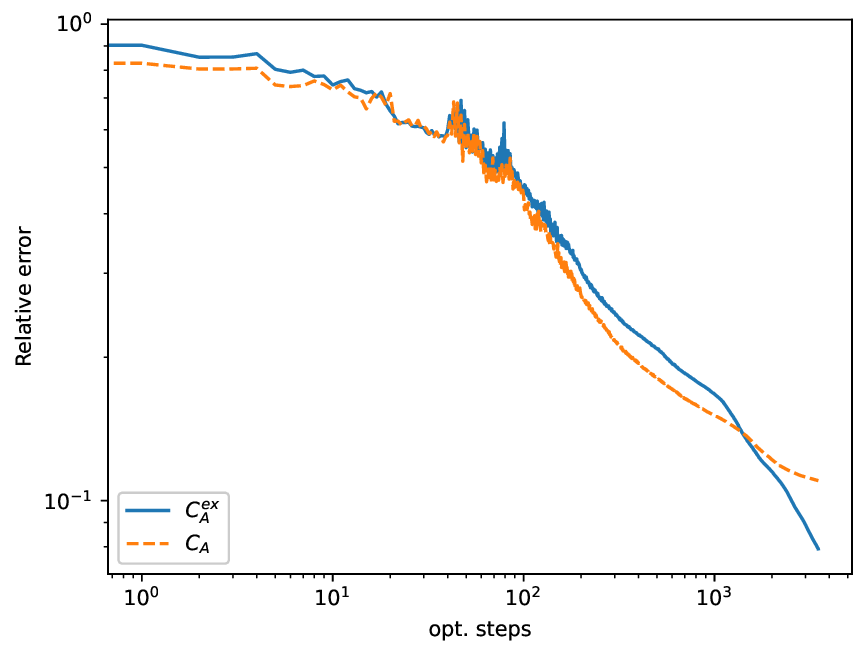}} 
    \hfill
    \subfloat[$m = 5$ ]{\includegraphics[width=0.5\columnwidth]{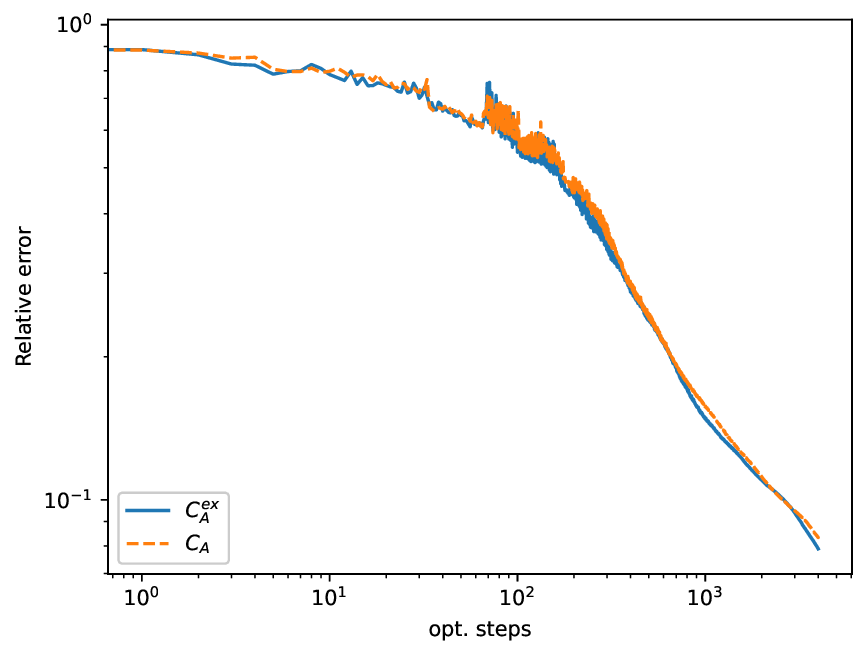}}
    
     \subfloat[$m = 6$]{\includegraphics[width=0.5\columnwidth]{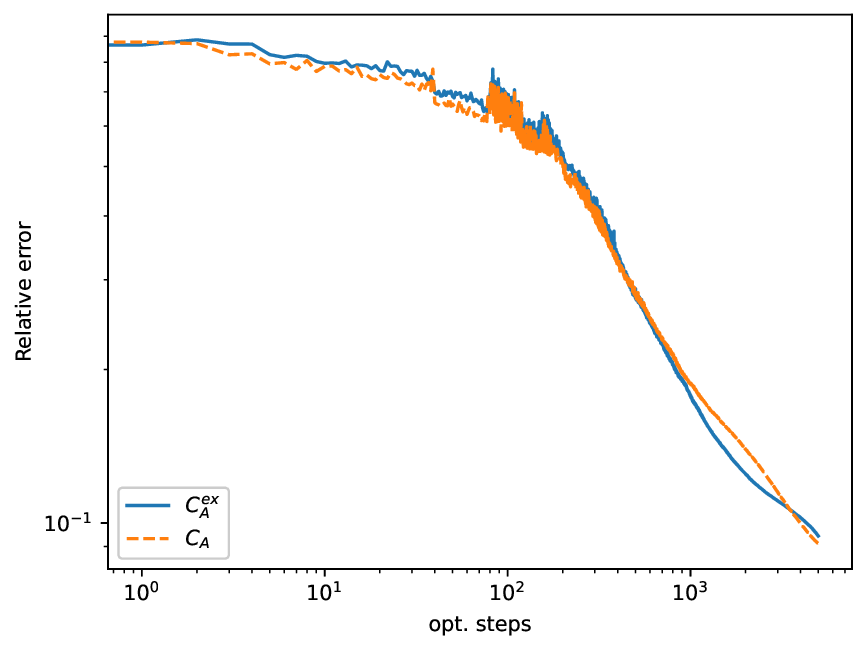}}
     \hfill
    \subfloat[$m = 10$]{\includegraphics[width=0.5\columnwidth]{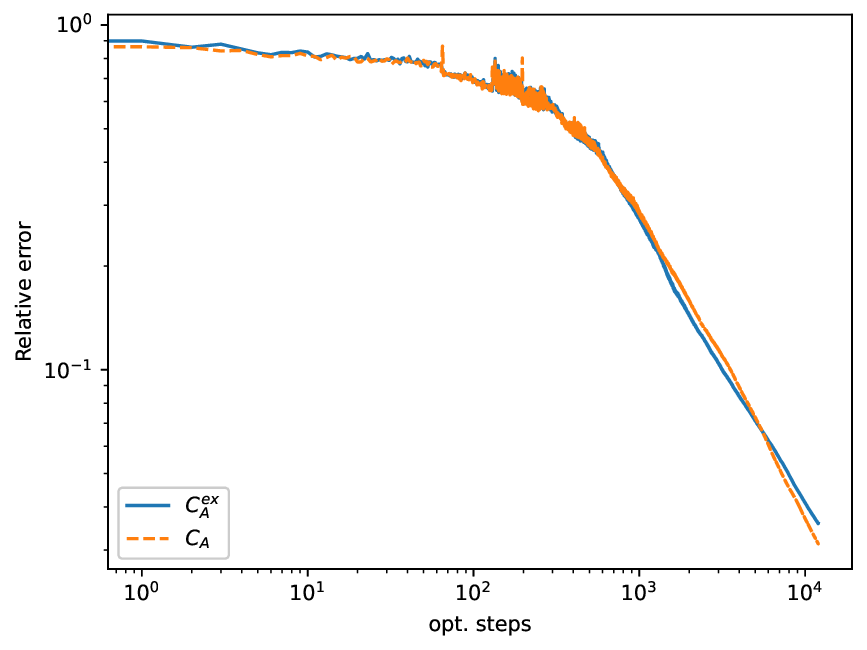}}     
\caption{Results of variational optimization of NN and NNN Heisenberg model using parameterized quantum circuits with and without NNN gates. The vertical axis is the relative error away from exact lowest energy, while the horizontal axis is the number of function evaluations. We trial only circuits $C_A$ with only NN gate-$A$ gates---shown as the dashed lines--- and $C_A^{\mathrm{ex}}$ with both NN and NNN gates---shown as solid lines. Exact ground energy $E_0 = -14.7262$ for a lattice with $8$ sites. We considered four different circuit layers, $m=3,5,6,10$. No significant relative advantage in using circuits with additional NNN gates is seen.}  
\label{Fig:Heisenberg_long}
\end{figure}

The results of learning ground states with the two circuits for various number of circuit layers are presented in Fig.~\ref{Fig:Heisenberg_long}. It can be seen that, though the Hamiltonian includes both NN and NNN interactions, there is no relative advantage in simulating it with circuits consisting of both NN and non-NN gates. In fact, in some cases, circuits with only NN gates performed a little better than those with additional NNN gates. The poorer performance of $C_A^{\mathrm{ex}}$ over $C_A$ may be attributed to boundary effects, due to lack of periodic boundary condition.

\subsection{Fidelity results of learning random states}
In order to further test the particle-conserving circuits and gates, we look at the problem of learning quantum states in  Fock space $\mathcal{H}_{N,L}$. The states are uniformly sampled according to Haar measure. The simulations performed in this section are exact. We set the number of random samples $N_S = 250$, and the number of trials $N_T = 10$. The Fock spaces considered are: (a) $\mathcal{H}_{2,4}$, for $2$ particles on $4$ sites, (b) $\mathcal{H}_{3,5}$, for $3$ particles on $5$ sites, and (c) $\mathcal{H}_{3,6}$, for $3$ particles on $6$ sites. The number of classical optimization steps is set such that the ratio of the number of evaluations to the Fock space dimension scales roughly the same for all lattices. For $\mathcal{H}_{2,4}$, we set this number equals to $1000$. We first test the learning capacity of the particle-conserving gates through circuits made up of only NN gates, and later we compare with circuits that have additional long-range gates.

\begin{figure}
    \centering
    \subfloat[$2$ particles on $4$ sites, $\mathcal{H}_{2,4}$]{\includegraphics[width=0.5\columnwidth]{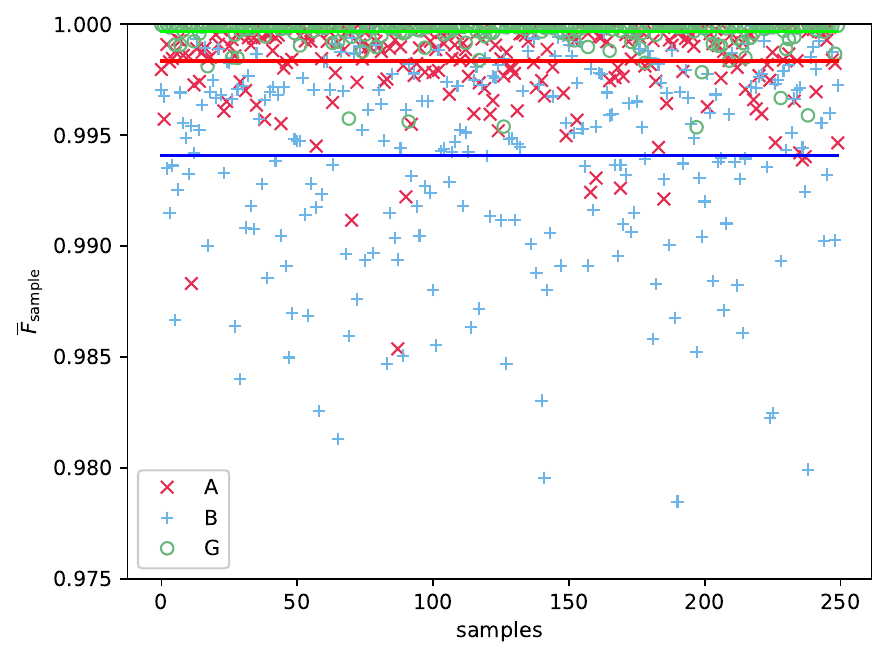}} 
    \hfill
    \subfloat[$3$ particles on $5$ sites, $\mathcal{H}_{3,5}$]{\includegraphics[width=0.5\columnwidth]{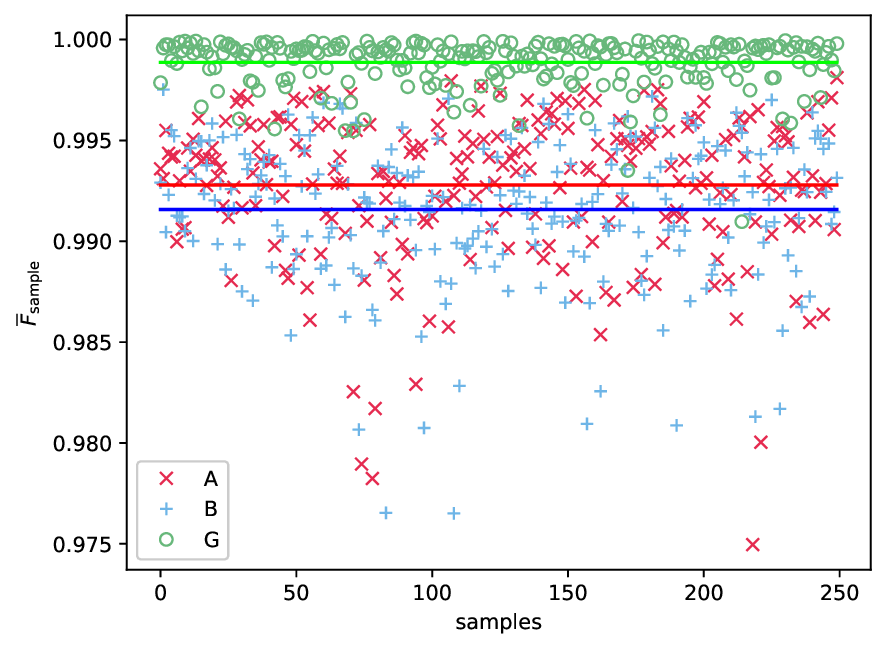}}
    
    \subfloat[$3$ particles on $6$ sites, $\mathcal{H}_{3,6}$]{\includegraphics[width=\columnwidth]{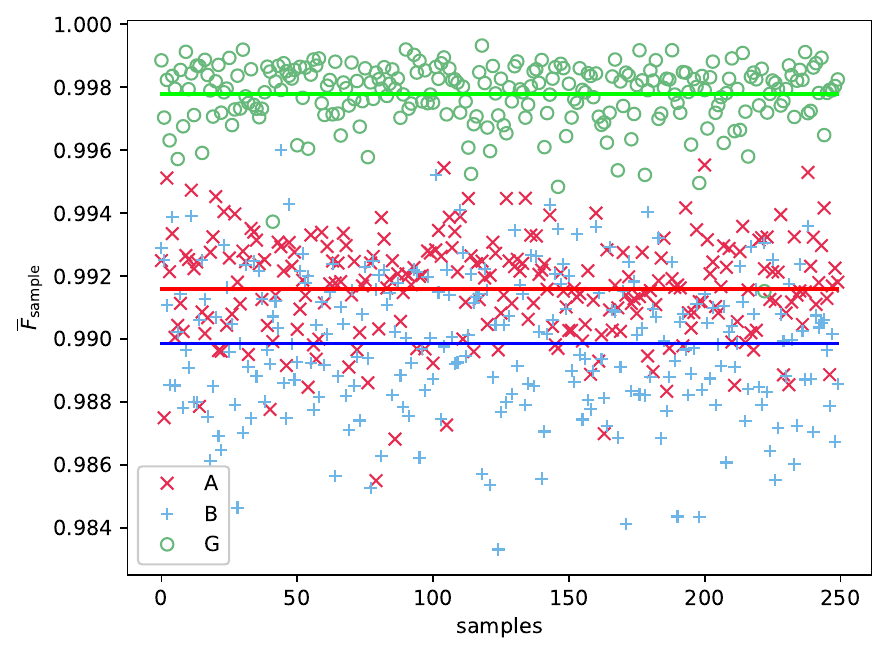}}
\caption{Plots of average fidelity $\bar{F}_{\mathrm{sample}}$ (i.e. fidelity per trial) of learning uniform randomly sampled states in Fock space $\mathcal{H}_{N,L}$ with brick-wall circuits $C_A$, $C_B$, and $C_G$. The Fock spaces considered are (a) $\mathcal{H}_{2,4}$ (b) $\mathcal{H}_{3,5}$, and (c) $\mathcal{H}_{3,6}$. The data points are the average fidelity $\bar{F}_{\mathrm{sample}}$, averaged over $N_T = 10$ trials, and plotted against the $N_S = 250$ random samples. The lines are the means $f$ (i.e. average fidelity per sample per trial) for the corresponding data points. Overall, gate $G$ has superior learning capability than gate $A$, which in turn is also better than gate $B$.}  
\label{Fig:Learning with ABG gates}
\end{figure}

The average fidelity $\mathrm{\bar{F}}_{sample}$ (i.e. fidelity per trial) of learning in the sample spaces are presented in Fig.~\ref{Fig:Learning with ABG gates}. It is clear from the plots that for the examples considered, the circuit with the generalized particle-conserving gate $G$ has a superior learning capability than with gate $A$, which in turn also has a higher learning capability than gate $B$. 

To explain where the superior capability of the gate $G$ comes from, we examine the case of $\mathcal{H}_{2,4}$ in detail. The dimension of $\mathcal{H}_{2,4}$ is $d_{2,4} = \mathrm{dim}(\mathcal{H}_{2,4} ) = 6$, which means any element of the space $\mathcal{H}_{2,4}$ has $12$ real numbers. Therefore, for the circuits used in learning from $\mathcal{H}_{2,4}$, we set the number of layers $\# l =2$, so that both circuits $C_A$ and $C_B$ have $12$ free parameters, while circuit $C_G$ will have $24$ free parameters. The superior learning capability of gate $G$ comes from having more parameters than gates $A$ and $B$. The learning capability of circuits with gates $A$ and $B$ can be increased if we increase the number of parameters by increasing the depth. We examine this by increasing the depth of $C_A$ and $C_B$ by one and two layers. This will correspondingly increase the number of parameters to $\# p = 18$ and $24$ respectively. In our simulations, the parameters of the circuits, both extended and unextended, were initialized from the same point of the parameter space. By increasing the number of parameters of circuits with both $A$- and $B$-gates, the accuracy of learning increased as shown in Table~\ref{Table:RelativeError}. When the size of the original circuits $C_A$ and $C_B$ is doubled, so that the circuits now have $4$ layers of gates and a total of $24$ free parameters, the error of learning dropped to $1.443 \times 10^{-6}$ and $2.782 \times 10^{-4}$ respectively. To compare, the error of learning with circuit $C_G$ is $3.434 \times 10^{-4}$, but the circuit similarly has $24$ free parameters but only $2$ layers. Since it is known that deep circuits can be plagued with barren plateaus,\cite{cerezo2021cost, wang2021noise, mcclean2018barren,ragone2023unified} this therefore makes circuit $C_G$ better than circuits $C_A$ and $C_B$, as it is shallower. 

\begin{figure}
    \centering
    \subfloat[With gate $A$]{\includegraphics[width=0.5\columnwidth]{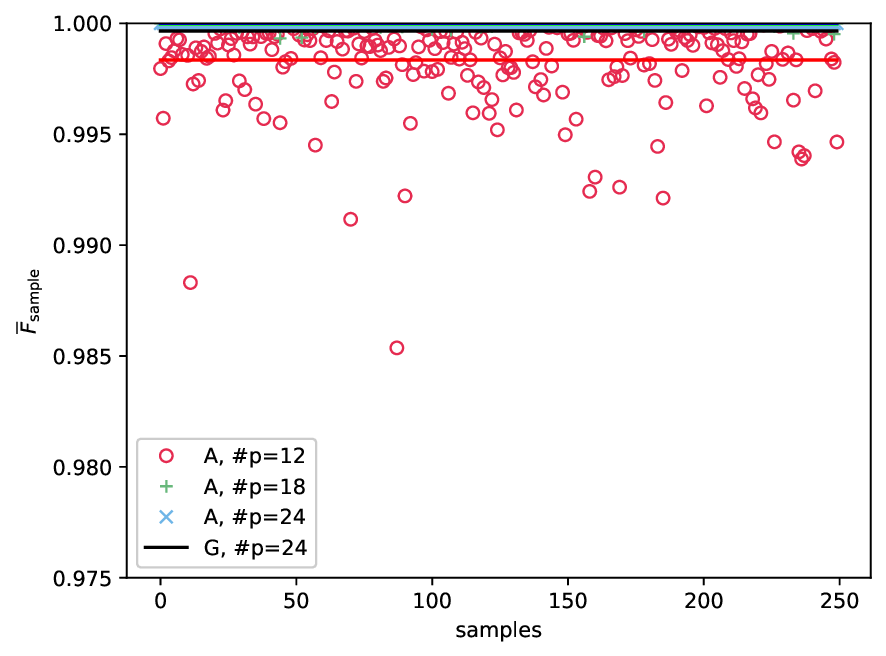}} 
    \hfill
    \subfloat[With gate $B$]{\includegraphics[width=0.5\columnwidth]{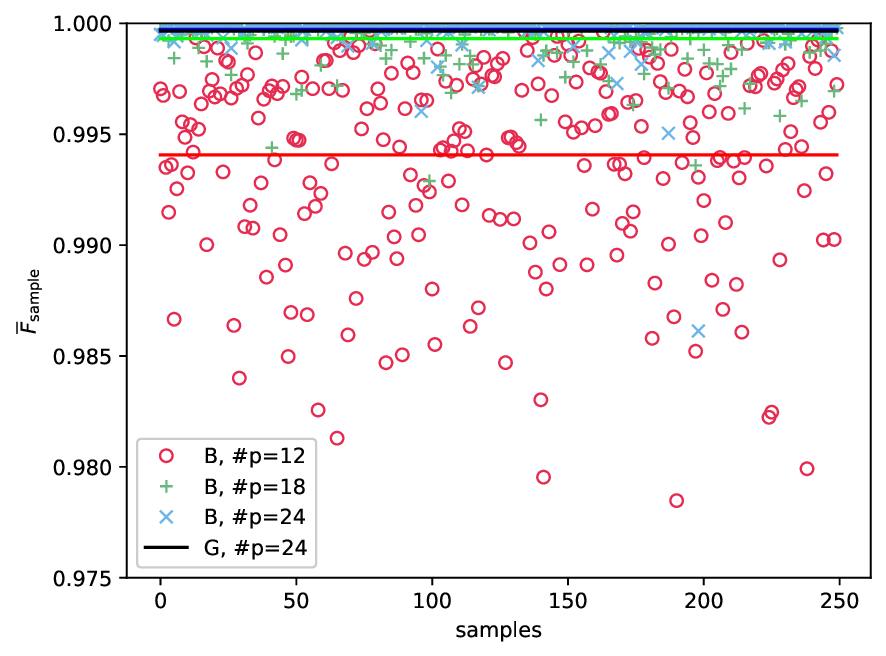}}
    
\caption{Learning random states in $\mathcal{H}_{2,4}$ with circuits $C_A$ and $C_B$ with increased number of parameters, $\# p =12,18, 24$. The accuracy of both gates $A$ and $B$ increased with the increased number of parameters. The data points of the fidelity per trial, $\bar{F}_{\mathrm{sample}}$, for the number of parameters $\# p=18,24$ all nearly approached the value of $1$ and therefore not quite visible, especially for gate $A$. The lines are the average fidelity (for all samples and trials) for the three different number of parameters considered. The average fidelities in both cases of gates $A$ and $B$ are compared with that of gate $G$. The error in the average fidelity for the higher number of parameters is not greater than $\sim10^{-4}$ (see Table~\ref{Table:RelativeError}).}  
\label{Fig:Increased parameter space with gates A and B}
\end{figure}

\begin{table}
\begin{tabularx}{0.8\columnwidth} { 
  | >{\centering\arraybackslash}X 
  | >{\centering\arraybackslash}X 
  | >{\centering\arraybackslash}X | }
 \hline
\#p / gate & A & B \\
 \hline
 $12$ & $1.653 \times 10^{-3}$ & $5.927 \times 10^{-3}$ \\
 \hline  
 $18$  & $3.053 \times 10^{-5}$  & $6.864 \times 10^{-4}$  \\
\hline
 $24$  & $1.443 \times 10^{-6}$ & $2.782 \times 10^{-4}$ \\
 \hline
\end{tabularx}
\caption{Table shows the relative error in the average fidelity of $\mathcal{H}_{2,4}$ with number of parameters $\# p = 12,18, 24$ for gates $A$ and $B$. Increasing the number of parameters increased the accuracy of learning, although at the expense of increasing the depth of the circuit.}
\label{Table:RelativeError}
\end{table}

\begin{figure}
    \centering
    \subfloat[$m=3$ layers]{\includegraphics[width=0.5\columnwidth]{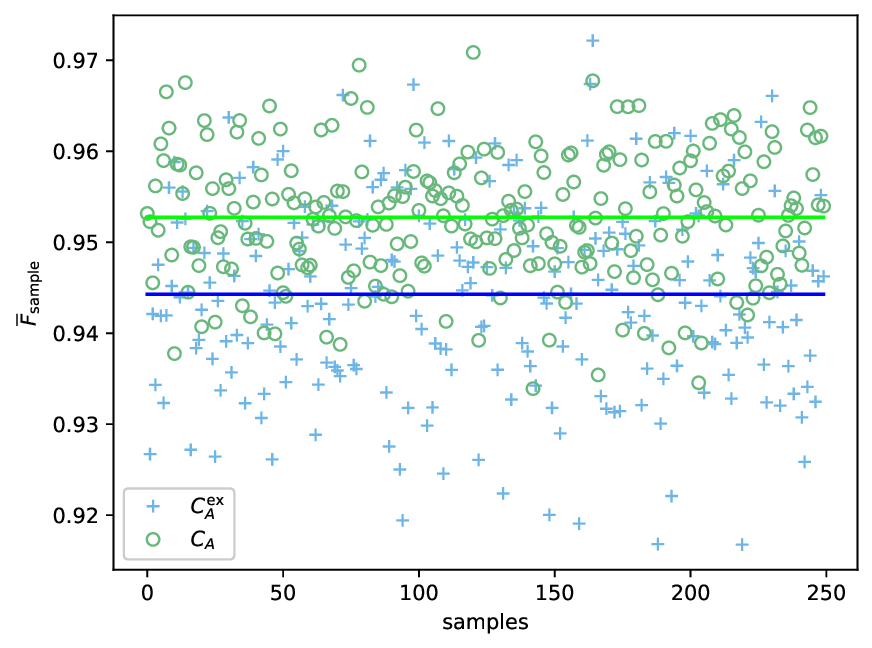}} 
    \hfill
    \subfloat[$m=4$ layers]{\includegraphics[width=0.5\columnwidth]{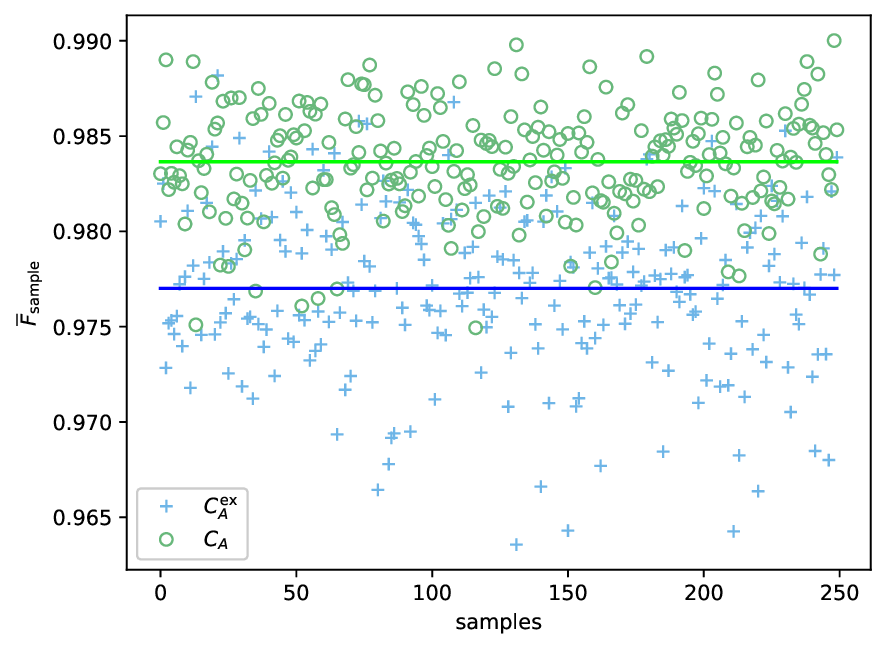}}
    
\caption{We compare circuits $C_A$ and $C_A^{\mathrm{ex}}$ for learning randomly sampled states in $\mathcal{H}_{3,6}$. The data points are the average fidelity per trial for the samples. The lines are the average fidelity per trial per sample corresponding to the data points.}  
\label{Fig:Short versus long-range circuits}
\end{figure}

Finally, we again compare the efficiency of circuits with and without long-range gates. We consider the learning of randomly sampled states in the Fock space $\mathcal{H}_{3,6}$, with circuit $C_A$ with only NN gates and circuit $C_A^{\mathrm{ex}}$ with both NN and NNN gates (e.g. see Fig.~\ref{Fig:Circuit Example with NN and NNN gates}). The results in Fig.~\ref{Fig:Short versus long-range circuits} shows that circuit $C_A$ with only NN gates performed better than circuit $C_A^{\mathrm{ex}}$. This conclusion remains the same for the other example Fock spaces we considered. We expect this to remain true for other particle-conserving gate types. Similar to the case of Heisenberg Hamiltonian simulations, the poorer performance of $C_A^{\mathrm{ex}}$ over $C_A$ may be attributed to lack of periodic boundary condition.

\section{Conclusion}
\label{Sec:Conclusion}
In this paper, we imported some ideas from symmetric tensor networks to construct efficient particle-conserving gates. We considered only single-excitation gates on nearest-neighbouring and next-nearest-neighbouring qubits. We derived three different types of parameterizations for particle-conserving gates, namely, two special gates with two parameters, and a third generalized particle-conserving gate $G$ with only four parameters. We then used the gates as ``modules'' to construct particle-conserving brick-wall parameterized quantum circuits. We presented algorithms for the construction of brick-wall circuits with either only nearest-neighbouring (NN) or both NN and non-NN gates, with open boundary condition. 

We tested the circuits in three case scenarios. In the first, we considered learning ground state of Heisenberg spin chain with only NN interactions. For this, we utilized circuits with only NN gates for the optimization, and compared the performance of the particle-conserving gates. In all experiments, we found the generalized particle-conserving gate $G$ to have the best performance. In the second scenario, we looked at learning randomly sampled quantum states. We again found gate $G$ to be the best of all the particle-conserving gate types. Though we found that the accuracy of circuits with the other gate types can be improved by increasing the circuit depth, but in general these circuits would become plagued with barren plateaus. Thirdly, we compare the learning capacity of circuits with and without non-NN gates. Specifically, we compare two circuits, one with only NN gates and another with both NN and next-NN gates. We used both circuits to learn the ground state of Heisenberg spin chain with next-nearest interaction terms. We found that circuits with both NN and NNN gates did not perform better than circuits with only NN gates, though this may be attributed to lack of periodic boundary condition.

\section{Code availability}
The codes used for this research can be made available upon reasonable request.

\section{Acknowledgement}
I thank Javier Osca Cotarelo for his preliminary code implementation in Qiskit during the initial stage of this project. I also thank Jiri Vala for valuable suggestions that improved the paper. This work is supported with funding from Enterprise Ireland’s DTIF programme of the Department of Business, Enterprise, and Innovation, project QCoIr Quantum Computing in Ireland: A Software Platform for Multiple Qubit Technologies No. DT 2019 0090B.

\appendix

\section{Mapping Heisenberg XXZ Model to Hardcore Bose-Hubbard Model}
\label{App:XXZ2BHM}
The Heisenberg XXZ model can be mapped to the hardcore Bose-Hubbard (HCBH) model using the following 
\begin{align}
\label{Eq:Pauli Second-quantized maps}
    X_i & = a_i + a^{\dagger}_i, \nonumber \\
    Y_i & = -i(a_i - a^{\dagger}_i),  \\
    Z_i & = 1 - 2n_i \nonumber,
\end{align}
where $a_i$($a^{\dagger}_i$) is a hardcore annihilation (creation) operator. The operators satisfy the following relations:
\begin{align}
    & \left[ a_i, a_j^{\dagger} \right] = \delta_{ij}, \\
    & a_i^2 = \left(a_i^{\dagger} \right)^2 = 0, \\
    & n_i = a_i^{\dagger} a_i,
\end{align}
where the second relation imposes the hardcore constraint---no site can have more than one particle. With a minor algebra, the XXZ Hamiltonian in Eq.~\eqref{Eq:XXZModel} can be written, up to some irrelevant terms, as 
\begin{equation}
\label{Eq: Bose-Hubbard model}
    H_{\mathrm{HCBH}} = \sum_i \left( a_i^{\dagger} a_{i+1} + a_{1+1}^{\dagger}a_i + \Delta n_i n_{1+1} \right),
\end{equation}
where $\Delta = 2\gamma$. 

This Hamiltonian $H_{\mathrm{HCBH}}$ commutes with the total particle number operator $N = \sum n_i$. Therefore, the total particle number is conserved. Furthermore, the particle number conservation of this Hamiltonian is related to the conservation of magnetization of the XXZ model. Indeed, using the third map in Eq.~\eqref{Eq:Pauli Second-quantized maps}, $M = \sum_i Z_i = L-2N$. Therefore, if $N$ is conserved, so is $M$. In particular, half-filling of the HCBH corresponds to the zero magnetization of the XXZ model. Therefore, particle-conserving circuits can also be used to simulate spin models.

\section{Generality of the particle-conserving gate G}
\label{Sec:ProofofGeneralPCgate}
We now prove that the particle-conserving gate $G(\alpha, \theta, \phi_1, \phi_2)$ given in Sec.~\ref{Sec:Generalized PC gate} is indeed general. 

Let the general particle-conserving two-qubit unitary be
\begin{equation}
U  = 
\begin{pmatrix}
e^{i \omega_1} & 0 & 0 & 0 \\
0 &  e^{i \alpha} e^{-i \frac{\phi_1+\phi_2}{2}} \cos{\theta}  & -e^{i \alpha} e^{-i\frac{\phi_1 - \phi_2}{2}} \sin \theta  & 0  \\
0  &  e^{i \alpha} e^{i\frac{\phi_1 - \phi_2}{2}} \sin \theta  & e^{i \alpha} e^{i\frac{\phi_1+\phi_2}{2}} \cos \theta & 0 \\
0 & 0  & 0 & 0  & e^{i \omega_2}
\end{pmatrix},
\end{equation}
where all parameters are real. Using tensor product of two single qubit unitaries 
\begin{equation}
T = 
\begin{pmatrix}
e^{-i \omega_1/2} & \\
& e^{-i \omega_2/2}
\end{pmatrix}
\otimes 
\begin{pmatrix}
e^{-i \omega_1/2} & \\
& e^{-i \omega_2/2}
\end{pmatrix},
\end{equation}
$U$ can be transformed to $U'$ using  $U' = T U = UT$ 
\begin{equation}
U'  = 
\begin{pmatrix}
1  &  0 & 0 & 0  \\
0  &  e^{i \alpha'}e^{-i\frac{\phi_1+\phi_2}{2}}\cos{\theta}   & -e^{i \alpha'}e^{-i\frac{\phi_1 - \phi_2}{2}}\sin{\theta}  &  0 \\
0  &  e^{i \alpha'}e^{i\frac{\phi_1 - \phi_2}{2}} \sin{\theta}   & e^{i \alpha'}e^{i\frac{\phi_1+\phi_2}{2}} \cos{\theta} & 0 \\
0 & 0  & 0 &  0 & 1
\end{pmatrix},
\end{equation}
where $\alpha' = \alpha - \frac{\omega_1 + \omega_2}{2}$ is once again a real parameter. Therefore, up to a product of two single-qubit unitaries, matrix $U'$ is the general particle-conserving two-qubit unitary, which is the form of $G$ given in the main text.

\bibliographystyle{plain}
\bibliography{References}

\begin{thebibliography}{10}

\bibitem{anselmetti2021local}
Gian-Luca~R Anselmetti, David Wierichs, Christian Gogolin, and Robert~M
  Parrish.
\newblock Local, expressive, quantum-number-preserving vqe ans{\"a}tze for
  fermionic systems.
\newblock {\em New Journal of Physics}, 23(11):113010, 2021.

\bibitem{arrazola2022universal}
Juan~Miguel Arrazola, Olivia Di~Matteo, Nicol{\'a}s Quesada, Soran Jahangiri,
  Alain Delgado, and Nathan Killoran.
\newblock Universal quantum circuits for quantum chemistry.
\newblock {\em Quantum}, 6:742, 2022.

\bibitem{ayeni2016simulation}
Babatunde~M Ayeni, Sukhwinder Singh, Robert~NC Pfeifer, and Gavin~K Brennen.
\newblock Simulation of braiding anyons using matrix product states.
\newblock {\em Physical Review B}, 93(16):165128, 2016.

\bibitem{barkoutsos2018quantum}
Panagiotis~Kl Barkoutsos, Jerome~F Gonthier, Igor Sokolov, Nikolaj Moll, Gian
  Salis, Andreas Fuhrer, Marc Ganzhorn, Daniel~J Egger, Matthias Troyer,
  Antonio Mezzacapo, et~al.
\newblock Quantum algorithms for electronic structure calculations:
  Particle-hole hamiltonian and optimized wave-function expansions.
\newblock {\em Physical Review A}, 98(2):022322, 2018.

\bibitem{bauer2011implementing}
Bela Bauer, Philippe Corboz, Rom{\'a}n Or{\'u}s, and Matthias Troyer.
\newblock Implementing global abelian symmetries in projected entangled-pair
  state algorithms.
\newblock {\em Physical Review B}, 83(12):125106, 2011.

\bibitem{bharti2022noisy}
Kishor Bharti, Alba Cervera-Lierta, Thi~Ha Kyaw, Tobias Haug, Sumner
  Alperin-Lea, Abhinav Anand, Matthias Degroote, Hermanni Heimonen, Jakob~S
  Kottmann, Tim Menke, et~al.
\newblock Noisy intermediate-scale quantum algorithms.
\newblock {\em Reviews of Modern Physics}, 94(1):015004, 2022.

\bibitem{cerezo2021variational}
Marco Cerezo, Andrew Arrasmith, Ryan Babbush, Simon~C Benjamin, Suguru Endo,
  Keisuke Fujii, Jarrod~R McClean, Kosuke Mitarai, Xiao Yuan, Lukasz Cincio,
  et~al.
\newblock Variational quantum algorithms.
\newblock {\em Nature Reviews Physics}, 3(9):625--644, 2021.

\bibitem{cerezo2021cost}
Marco Cerezo, Akira Sone, Tyler Volkoff, Lukasz Cincio, and Patrick~J Coles.
\newblock Cost function dependent barren plateaus in shallow parametrized
  quantum circuits.
\newblock {\em Nature communications}, 12(1):1791, 2021.

\bibitem{gard2020efficient}
Bryan~T Gard, Linghua Zhu, George~S Barron, Nicholas~J Mayhall, Sophia~E
  Economou, and Edwin Barnes.
\newblock Efficient symmetry-preserving state preparation circuits for the
  variational quantum eigensolver algorithm.
\newblock {\em npj Quantum Information}, 6(1):1--9, 2020.

\bibitem{grant2019initialization}
Edward Grant, Leonard Wossnig, Mateusz Ostaszewski, and Marcello Benedetti.
\newblock An initialization strategy for addressing barren plateaus in
  parametrized quantum circuits.
\newblock {\em Quantum}, 3:214, 2019.

\bibitem{konig2010anyonic}
Robert K{\"o}nig and Ersen Bilgin.
\newblock Anyonic entanglement renormalization.
\newblock {\em Physical Review B}, 82(12):125118, 2010.

\bibitem{lacroix2023symmetry}
Denis Lacroix, Edgar~Andres Ruiz~Guzman, and Pooja Siwach.
\newblock Symmetry breaking/symmetry preserving circuits and symmetry
  restoration on quantum computers: A quantum many-body perspective.
\newblock {\em The European Physical Journal A}, 59(1):3, 2023.

\bibitem{lee2018generalized}
Joonho Lee, William~J Huggins, Martin Head-Gordon, and K~Birgitta Whaley.
\newblock Generalized unitary coupled cluster wave functions for quantum
  computation.
\newblock {\em Journal of chemical theory and computation}, 15(1):311--324,
  2018.

\bibitem{mcclean2018barren}
Jarrod~R McClean, Sergio Boixo, Vadim~N Smelyanskiy, Ryan Babbush, and Hartmut
  Neven.
\newblock Barren plateaus in quantum neural network training landscapes.
\newblock {\em Nature communications}, 9(1):4812, 2018.

\bibitem{mcclean2016theory}
Jarrod~R McClean, Jonathan Romero, Ryan Babbush, and Al{\'a}n Aspuru-Guzik.
\newblock The theory of variational hybrid quantum-classical algorithms.
\newblock {\em New Journal of Physics}, 18(2):023023, 2016.

\bibitem{mcculloch2002non}
Ian~P McCulloch and Mikl{\'o}s Gul{\'a}csi.
\newblock The non-abelian density matrix renormalization group algorithm.
\newblock {\em Europhysics Letters}, 57(6):852, 2002.

\bibitem{Note1}
We defined a symmetry basis as the basis where an operator may become block
  diagonal, where each block is identified with a specific charge. The states
  in each charge sector are referred to as \protect \emph {degenerate states}
  of the sector. The word ``degenerate'' is used in the sense that measurement
  of the states in a particular charge sector all give the same charge outcome,
  hence degenerate.

\bibitem{Note2}
The idea behind the formula is that, since only $(L-1)$ number of NN two-qubit
  gates ``tiles'' a layer, then to have a circuit with $d_{N,L}$ number of
  gates, a total of $d_{N,L}/(L-1)$ layers will be needed. To avoid fractions,
  we round up. But this will increase the number of gates needed, and therefore
  the number of parameters. The extra parameters can be fixed to bring down the
  parameter count.

\bibitem{Note3}
Since circuits $C_A$ have more gates than $C_A^{\protect \mathrm {ex}}$, the
  former circuit will \protect \emph {a priori} have more parameters than the
  latter. To make the number of parameters in the two circuits the same, we
  adopt a convention that the last extra parameters in $C_A$ are fixed all
  equal.

\bibitem{patti2021entanglement}
Taylor~L Patti, Khadijeh Najafi, Xun Gao, and Susanne~F Yelin.
\newblock Entanglement devised barren plateau mitigation.
\newblock {\em Physical Review Research}, 3(3):033090, 2021.

\bibitem{peruzzo2014variational}
Alberto Peruzzo, Jarrod McClean, Peter Shadbolt, Man-Hong Yung, Xiao-Qi Zhou,
  Peter~J Love, Al{\'a}n Aspuru-Guzik, and Jeremy~L O’brien.
\newblock A variational eigenvalue solver on a photonic quantum processor.
\newblock {\em Nature communications}, 5(1):4213, 2014.

\bibitem{pfeifer2010simulation}
Robert~NC Pfeifer, Philippe Corboz, Oliver Buerschaper, Miguel Aguado, Matthias
  Troyer, and Guifre Vidal.
\newblock Simulation of anyons with tensor network algorithms.
\newblock {\em Physical Review B}, 82(11):115126, 2010.

\bibitem{pfeifer2015finite}
Robert~NC Pfeifer and Sukhwinder Singh.
\newblock Finite density matrix renormalization group algorithm for anyonic
  systems.
\newblock {\em Physical Review B}, 92(11):115135, 2015.

\bibitem{preskill2018quantum}
John Preskill.
\newblock Quantum computing in the nisq era and beyond.
\newblock {\em Quantum}, 2:79, 2018.

\bibitem{Qiskit}
{Qiskit contributors}.
\newblock Qiskit: An open-source framework for quantum computing, 2023.

\bibitem{ragone2023unified}
Michael Ragone, Bojko~N Bakalov, Fr{\'e}d{\'e}ric Sauvage, Alexander~F Kemper,
  Carlos~Ortiz Marrero, Martin Larocca, and M~Cerezo.
\newblock A unified theory of barren plateaus for deep parametrized quantum
  circuits.
\newblock {\em arXiv preprint arXiv:2309.09342}, 2023.

\bibitem{robertson2022escaping}
Niall~F Robertson, Albert Akhriev, Jiri Vala, and Sergiy Zhuk.
\newblock Escaping barren plateaus in approximate quantum compiling.
\newblock {\em arXiv preprint arXiv:2210.09191}, 2022.

\bibitem{robertson2023approximate}
Niall~F Robertson, Albert Akhriev, Jiri Vala, and Sergiy Zhuk.
\newblock Approximate quantum compiling for quantum simulation: A tensor
  network based approach.
\newblock {\em arXiv preprint arXiv:2301.08609}, 2023.

\bibitem{sack2022avoiding}
Stefan~H Sack, Raimel~A Medina, Alexios~A Michailidis, Richard Kueng, and
  Maksym Serbyn.
\newblock Avoiding barren plateaus using classical shadows.
\newblock {\em PRX Quantum}, 3(2):020365, 2022.

\bibitem{schmoll2020programming}
Philipp Schmoll, Sukhbinder Singh, Matteo Rizzi, and Rom{\'a}n Or{\'u}s.
\newblock A programming guide for tensor networks with global su (2) symmetry.
\newblock {\em Annals of Physics}, 419:168232, 2020.

\bibitem{singh2010tensor}
Sukhwinder Singh, Robert~NC Pfeifer, and Guifr{\'e} Vidal.
\newblock Tensor network decompositions in the presence of a global symmetry.
\newblock {\em Physical Review A}, 82(5):050301, 2010.

\bibitem{singh2011tensor}
Sukhwinder Singh, Robert~NC Pfeifer, and Guifre Vidal.
\newblock Tensor network states and algorithms in the presence of a global u(1)
  symmetry.
\newblock {\em Physical Review B}, 83(11):115125, 2011.

\bibitem{singh2014matrix}
Sukhwinder Singh, Robert~NC Pfeifer, Guifre Vidal, and Gavin~K Brennen.
\newblock Matrix product states for anyonic systems and efficient simulation of
  dynamics.
\newblock {\em Physical Review B}, 89(7):075112, 2014.

\bibitem{singh2012tensor}
Sukhwinder Singh and Guifre Vidal.
\newblock Tensor network states and algorithms in the presence of a global su
  (2) symmetry.
\newblock {\em Physical Review B}, 86(19):195114, 2012.

\bibitem{vatan2004optimal}
Farrokh Vatan and Colin Williams.
\newblock Optimal quantum circuits for general two-qubit gates.
\newblock {\em Physical Review A}, 69(3):032315, 2004.

\bibitem{wang2021noise}
Samson Wang, Enrico Fontana, Marco Cerezo, Kunal Sharma, Akira Sone, Lukasz
  Cincio, and Patrick~J Coles.
\newblock Noise-induced barren plateaus in variational quantum algorithms.
\newblock {\em Nature communications}, 12(1):6961, 2021.

\bibitem{wang2023trainability}
Yabo Wang, Bo~Qi, Chris Ferrie, and Daoyi Dong.
\newblock Trainability enhancement of parameterized quantum circuits via
  reduced-domain parameter initialization.
\newblock {\em arXiv preprint arXiv:2302.06858}, 2023.

\bibitem{weichselbaum2012non}
Andreas Weichselbaum.
\newblock Non-abelian symmetries in tensor networks: A quantum symmetry space
  approach.
\newblock {\em Annals of Physics}, 327(12):2972--3047, 2012.

\bibitem{xia2020qubit}
Rongxin Xia and Sabre Kais.
\newblock Qubit coupled cluster singles and doubles variational quantum
  eigensolver ansatz for electronic structure calculations.
\newblock {\em Quantum Science and Technology}, 6(1):015001, 2020.

\bibitem{xie2022qubit}
Qing-Xing Xie, Wen-gang Zhang, Xu-Sheng Xu, Sheng Liu, and Yan Zhao.
\newblock Qubit unitary coupled cluster with generalized single and paired
  double excitations ansatz for variational quantum eigensolver.
\newblock {\em International Journal of Quantum Chemistry}, 122(24):e27001,
  2022.

\bibitem{yordanov2020efficient}
Yordan~S Yordanov, David~RM Arvidsson-Shukur, and Crispin~HW Barnes.
\newblock Efficient quantum circuits for quantum computational chemistry.
\newblock {\em Physical Review A}, 102(6):062612, 2020.

\bibitem{zhang2004minimum}
Jun Zhang, Jiri Vala, Shankar Sastry, and K~Birgitta Whaley.
\newblock Minimum construction of two-qubit quantum operations.
\newblock {\em Physical review letters}, 93(2):020502, 2004.

\bibitem{zhang2022escaping}
Kaining Zhang, Liu Liu, Min-Hsiu Hsieh, and Dacheng Tao.
\newblock Escaping from the barren plateau via gaussian initializations in deep
  variational quantum circuits.
\newblock {\em Advances in Neural Information Processing Systems},
  35:18612--18627, 2022.

\end{thebibliography}

\end{document}